\documentclass[fleqn,usenatbib]{mnras}

\usepackage{newtxtext,newtxmath}

\usepackage[T1]{fontenc}
\usepackage{ae,aecompl}

\usepackage{graphicx}	
\usepackage{amsmath}	
\usepackage[colorinlistoftodos]{todonotes}
\usepackage{hyperref}
\hypersetup{
    colorlinks=true,
    linkcolor=red,
    filecolor=magenta,
    urlcolor=cyan,
}

\usepackage{caption}
\usepackage{parskip}
\usepackage{float}
\usepackage{placeins}
\usepackage{subcaption}

\newcommand{\ml}[1]{{\textcolor{black}{#1}}}
\newcommand{\ks}[1]{{\textcolor{black}{#1}}}
\newcommand{\ksn}[1]{{\textcolor{black}{#1}}}
\newcommand{\ksr}[1]{{\textcolor{black}{#1}}}
\newcommand{\mln}[1]{{\textcolor{black}{#1}}}
\newcommand{\ksd}[1]{{\textcolor{black}{#1}}}
\newcommand{\kss}[1]{{\textcolor{black}{#1}}}
\newcommand{\ksnn}[1]{{\textcolor{black}{#1}}}
\newcommand{\ksfour}[1]{{\textcolor{black}{#1}}}
\newcommand{\ksfive}[1]{{\textcolor{black}{#1}}}

\title[Classifying Multiwavelength Transients with ML]{Classification of Multiwavelength Transients with Machine Learning}

\author[K. Sooknunan et al.]{
K. Sooknunan$^{1,2}$,
M. Lochner$^{3,4,5}$,
Bruce  A. Bassett$^{1,4,5,6}$,
H. V.  Peiris$^{7,8}$,
R. Fender$^{9,11}$,
\newauthor 
\;A. J. Stewart$^{10,9}$,
M. Pietka$^{9}$,
P. A. Woudt$^{11}$,
J. D. McEwen$^{12}$,
O. Lahav$^{7}$
\\~\\
$^{1}$Department of Maths and Applied Maths, University of Cape Town, Cape Town, South Africa\\
$^{2}$Astrophysics Group, Imperial College London, Blackett Laboratory, Prince Consort Road, London SW7 2AZ, UK\\
$^{3}$Department of Physics and Astronomy, University of the Western Cape, Bellville, Cape Town, 7535, South Africa\\
$^{4}$South African Radio Astronomy Observatory (SARAO), The Park, Park Road, Pinelands, Cape Town 7405, South Africa\\
$^{5}$African Institute for Mathematical Sciences, 6 Melrose Road, Muizenberg, 7945, South Africa\\
$^{6}$ South African Astronomical Observatory, Observatory, Cape Town, 7925, South Africa\\ 
$^{7}$Department of Physics and Astronomy, University College London, Gower Street, London WC1E 6BT, UK\\
$^{8}$The Oskar Klein Centre for Cosmoparticle Physics, Department of Physics, Stockholm University, AlbaNova, 10691 Stockholm, Sweden\\
$^{9}$Astrophysics, Department of Physics, University of Oxford, Keble Road, Oxford OX1 3RH, UK\\
$^{10}$Sydney Institute for Astronomy, School of Physics, The University of Sydney, NSW 2006, Australia \\
$^{11}$Department of Astronomy, University of Cape Town, Cape Town, South Africa\\
$^{12}$Mullard Space Science Laboratory, University College London, Surrey RH5 6NT, UK
}

\date{Accepted XXX. Received YYY; in original form ZZZ}

\pubyear{2020}

\begin{document}
\label{firstpage}
\pagerange{\pageref{firstpage}--\pageref{lastpage}}
\maketitle

\begin{abstract}

With the advent of powerful telescopes such as the Square Kilometer Array and  the Vera C. Rubin Observatory, we are entering an era of multiwavelength transient astronomy that will lead to a dramatic increase in data volume.
\mln{Machine learning techniques are well suited to address this data challenge and rapidly classify newly detected transients.}
We present a multiwavelength classification algorithm consisting of three steps: (1) interpolation and augmentation of the data using Gaussian processes; (2) feature extraction using wavelets; (3) classification with random forests. Augmentation provides improved performance at test time by balancing the classes and adding diversity into the training set. \mln{In the first application of machine learning to the classification of real radio transient data, we apply our technique to the Green Bank Interferometer and other radio light curves.} \mln{We find }we are able to accurately classify most of the eleven classes of radio variables and transients after just eight hours of observations, achieving an overall test accuracy of \ksnn{78\%}. \mln{We fully investigate the impact of the small sample size of 82 publicly available light curves and use data augmentation techniques to mitigate the effect.} We \mln{also} show that on a significantly larger simulated representative training set that the algorithm achieves an overall accuracy of \ksnn{97\%}, illustrating that the method is likely to provide excellent performance on future surveys. Finally, we demonstrate the effectiveness of simultaneous multiwavelength observations by showing how incorporating just one optical data point into the analysis improves the \ksnn{accuracy of the worst performing class by 19\%}.\\~\\


\end{abstract}



\graphicspath{{plots/}}


\section{Introduction}
\label{sec:intro}
In the coming years, radio astronomy will enter a new era of deep field surveys with \mln{new telescopes such as} the Square Kilometre Array\footnote{www.skatelescope.org} (SKA) and its precursors, MeerKAT\footnote{www.ska.ac.za/science-engineering/meerkat} and the Australian Square Kilometre Array Pathfinder\footnote{www.atnf.csiro.au/projects/askap/index.html}  (ASKAP). These telescopes will achieve unprecedented sensitivity and resolution. Large science projects such as ThunderKAT \citep{Fender:2017}  \mln{on MeerKAT}  will dramatically increase the detected number of radio transients. In the past, radio transient datasets have been small, \ksn{allowing} spectroscopic classification of all objects of interest. As the \ksn{event rate} increases, follow-up resources must be prioritised by making use of early classification of the radio data. Machine learning algorithms have proven themselves invaluable in this context \citep{2010IJMPD..19.1049B}. 

There has been a substantial amount of work done with machine learning in astronomy over the last decade. This includes research done by \cite{BailerJones:2001ni} in stellar classification, image-based classification of supernovae \citep{Romano:2006,Bailey:2007} and classifying variable stars \citep{Richards:2011}. In recent years, these algorithms have been used successfully in classifying optical transients, such as classification of transients in SDSS images \citep{Buisson:2014}, supernovae (e.g. \cite{Newling:2011,Ishida:2013,Karpenka:2013,Lochner:2016,Moller:2019}), variable sources \citep{Farrell:2016} or general optical transients \citep{Mahabal2017}. Some machine learning methods have been investigated for the upcoming ASKAP survey for Variables and Slow Transients (VAST) \citep{Rebbapragada:2012,Murphy2013}, but these were only applied to optical data. 

In the burgeoning era of multimessenger astronomy, incorporating data from different telescopes could dramatically improve classification of events. A prime example of this is the MeerLICHT \footnote{www.meerlicht.uct.ac.za} telescope \citep{Bloemen:2016vqg}, an optical telescope \ksn{whose observing schedule is synchronised with that of the (night time) observations of the} radio telescope MeerKAT, resulting in simultaneous optical and radio observations of transients. \ksr{Indeed, a primary motivation for undertaking this work now is that MeerKAT has recently  begun  to  take  scientific  data  and  it  is  anticipated  to  detect hundreds  of  transients over the next few years}. Alert streams from telescopes such as Fermi\footnote{https://fermi.gsfc.nasa.gov} and the  Vera C. Rubin Observatory \footnote{www.lsst.org} \citep{LSST}	will also enable rapid coordination for multimessenger observations. Combining these data sources necessitates a new general framework for multimessenger machine learning.

In this paper we outline a method for the automatic classification of radio transients\ml{, based on \cite{Lochner:2016},} that makes use of multiwavelength data and machine learning. We define how to incorporate data in other wavelengths and alert streams from other telescopes. \ksr{The techniques proposed by \cite{Lochner:2016} were only used on optical supernovae. Here, we apply this technique to radio data and show that it can also be effective for objects that vary on dramatically different timescales; supernovae and flare stars, for instance, vary on the time scales of seconds, while AGN can vary on the time scale of months.}

\mln{Due to the limited quantity of available radio transient light curves, we develop a data augmentation technique to increase the sample size and improve classification performance. Our data augmentation technique achieves three things: it breaks up long light curves into smaller samples increasing the size of the training set, it balances the number of objects in each class which reduces bias in the machine learning and it adds a small amount of additional variance to the training data that is statistically consistent with the original light curve.}

We test our method on existing radio transient light curves, exploring the effects of non-representative training sets (i.e. when the algorithm is tested on objects that are dissimilar to those in the training set). We also demonstrate, with an example, the effect of including optical data to improve classification accuracy.


\section{Background}
\label{sec:back}


\subsection{Radio transients}

A transient is an astronomical object observed to have a time-dependent brightness (flux). The time scales of variability range from milliseconds to a few years. Radio transients have emission frequencies in the radio regime (for a review on radio transients see \cite{Fender:2011}). Transient events are typically divided into two kinds: incoherent synchrotron events and coherent burst events.

Incoherent synchrotron events are thought to be caused by a high energy phenomenon, from which a large amount of energy is released over a longer time scale (on the order of minutes or larger). When in a steady state, this energy release is limited to a brightness temperature of $  T \leq 10^{12} \mathrm{K}$ \citep{Fender:2015sca}.

Coherent bursts occur on much shorter time scales (on the order of seconds) and can have brightness temperatures up to $ 10^{30}\mathrm{K}$. This type of emission can only be observed using a special mode on radio telescopes \citep{Fender:2015sca}. For this study, therefore, we will only consider incoherent synchrotron transient events. We assume that a light curve is generated from measuring the fluxes from radio images. We include the following transients in our study: active galactic nuclei (AGN), algols, flare stars (FS), gamma ray bursts (GRBs), kilonovae, magnetars, novae, RS canum venticorums (RSCVn), supernovae (SNe), tidal disruption events (TDEs) and X-ray binaries (XRBs).

Looking at the raw radio light curves of transients, while some objects exhibit obvious differences (such as binary star systems and AGN) others look more similar. Contextual information (such as the location of the object) can be invaluable in telling the difference between classes, as can multiwavelength data. We begin by studying how well we can distinguish between classes using radio data alone, and then show how including contextual and multiwavelength information can improve the classification accuracy.

\subsection{Machine learning overview}


Machine learning can broadly be split into two approaches: supervised and unsupervised learning. In this study we use supervised learning. A supervised machine learning algorithm automatically learns a model given a set of known inputs and outputs, called a training set. The now trained algorithm can be given new inputs, called a test set, that it will then map to an output. In terms of a classification problem, the inputs constitute objects to be classified, and the outputs are the class labels assigned to each object. For an in depth review of machine learning see \cite{Mitchell:1997} or \cite{Mackay:2003}. 

\ks{In this study we used the ``random forests" algorithm \citep{RForests,Breiman:2001}. Random forests have been shown to outperform other algorithms in a variety of cases \citep{Caruana:2006,Liu:2013,Lochner:2016}. Deep learning algorithms\footnote{For a review on deep learning algorithms see \cite{Vargas:2018}.} such as LSTMs\footnote{A LSTM or Long Short Term Memory network is a type of deep learning algorithm that is primarily used when dealing with time series data. } \citep{Hochreiter:1997} would be an interesting approach to this problem. However we do not consider them here due to the requirement of large training sets and significant computational resources, especially in light of the excellent performance of faster traditional techniques.}

\subsubsection{Random Forests}

\ml{Ensemble methods like random forests \citep{RForests,Breiman:2001} build robust classifiers out of a multitude of weak learners such as decision trees.}
A decision tree creates a mapping, by making a series of ``yes/no" decisions, from an input vector (the feature vector) to an output label (the class) \citep{2010IJMPD..19.1049B}. The algorithm creates this mapping by making decisions based on whether or not a given component of the feature vector falls into some range.  One of the main drawbacks of decision trees is the high variance on the labels it outputs.

This problem can be overcome by training many separate trees and taking the average of the output. Random forests perform an additional step, each tree in the ``forest" is trained on a random subset of the total feature set. This leads to more robust overall predictions. We used the package \texttt{Scikit Learn}\footnote{www.scikit-learn.org} \citep{scikit-learn} to implement the random forest classifier.

\subsubsection{Feature extraction in machine learning}
Classical machine learning techniques can seldom use data in its raw format for classification. Feature extraction is a technique used to reduce the dimensionality of the data by summarizing the information contained in the original data. The features one uses should also be well-separated between classes. Taking a repeating light curve as an example, features one could extract are the frequency; the amplitude; the phase etc. For more on feature extraction see \cite{Li:2016}. \ml{An obvious choice of simple features for this problem would be changes of flux values over specific time periods. However we found these were inadequate to capture the variation between classes (see \textcolor{red}{Sec.} \ref{sec:simple}) and so we instead follow the feature extraction procedure used in \cite{Lochner:2016}.} In \textcolor{red}{Sec.}  \ref{sec:wavelet}, we outline the wavelet decomposition approach used, which resulted in much higher performance.

\subsubsection{Visualising features}
\label{sec:vis}

Visualising feature vectors is quite difficult because of their often high-dimensional nature. In this paper we use two methods for feature visualisation:

\textbf{$t$-distributed Stochastic Neighbor Embedding}\\

 One tool commonly used to visualise higher-dimensional spaces is $t$-distributed Stochastic Neighbor Embedding or $t$-SNE \citep{vanDerMaaten2008}. It works by computing the probability that two points are similar in the higher dimensional space based on its Euclidean distance. It does this for every pair of points in the feature set, then attempts to find a lower dimensional representation of these points that preserves the probability distribution. Thus points clustered in this lower dimensional representation correspond to points clustered in the original higher dimensional space. $t$-SNE uses Student's $t$-distribution when determining the degree of similarity of two points. \ksfour{It is important to note that distances are not preserved when transforming to the lower dimensional space but rather the probability distributions. Therefore, while these visualisations can be used to give us some intuition into how the data points are distributed in higher dimensional space, they can not be used to infer the classes of the data points.} We stress that $t$-SNE plots are useful tools for visualisation purposes only, and cannot be used as classifiers themselves due to their stochastic nature.\\~\\

\textbf{Uniform Manifold Approximation and Projection for Dimension Reduction}\\

\ksnn{Recently, a new visualisation and dimensionality reduction technique Uniform Manifold Approximation and Projection for Dimension Reduction (UMAP) from McInnes et al. (2018) has begun to overtake t-SNE in popularity, due to its superior computational performance and interpretability.}

\ksnn{We make use of both techniques to visualise our feature space.}

\subsubsection{Training, testing and cross validation}
In machine learning, a dataset is generally split into two main subsets: the training set and the test set. The algorithm is given the training set from which to learn the parameters of the model. The test set is reserved in order to check if the algorithm has learned an accurate model, and is only presented to the algorithm after the training step is complete. In some cases an algorithm can perform very well on the training set but perform poorly on the test set. This occurs when the algorithm overfits the model parameters to the test set and hence the model will not generalise to the training set. One method for overcoming this is known as $k$-fold cross validation. Instead of splitting the dataset into two subsets, the dataset is split into $k > 2$ subsets. One of these subsets is then used as the validation set while the others are used as the training set. This is then repeated $k$ times until all $k$ subsets has been used as a test set and average results are used \mln{to evaluate the algorithm}.

Machine learning models have two types of parameters. The first are the parameters that are learned by the algorithm during training as mentioned above. The second type is known as hyperparameters. These are parameters that are not learned during training but are set by the user. These parameters can also be optimised by specifying a range for each hyperparameter, searching through this hyperparameter space and \ml{using cross-validation to} choose the hyperparameters with the best performance. 
\kss{Three-fold cross validation was used to optimise the random forest algorithm hyperparameters.}

\subsubsection{Data Augmentation.}
\ksr{Data augmentation is widely used in machine learning when dealing with small datasets (\cite{Hoyle:2015,2019:Oviedo}). There are three main benefits of data augmentation: the first is the creation of larger training sets which is important for complex algorithms with many parameters (such as deep neural networks); the second benefit is that it helps avoid overfitting by introducing variation in the original dataset; finally it allows the balancing of classes which may otherwise cause biases in the classifier. Augmentation is used extensively in the field of deep learning for image classification: training images are rotated, cropped, translated and zoomed to create new examples which can dramatically improve final performance (e.g. \cite{2016:Kim,Perez:2017}). The technique Synthetic Minority Over-sampling Technique (SMOTE) is commonly used to balance class numbers and add variance by simulating new data in feature space (\cite{Chawla:2011}). \kss{\cite{Naul:2018} used data augmentation to help train a recurrent neural network in the classification of variable stars.} Recently, \cite{trotta:2018, Boone:2019} have made use of Gaussian processes to augment non-representative training data to increase accuracy for classification of supernovae and other optical transients.}

\subsubsection{Evaluating machine learning results}

The performance of a machine learning algorithm can be measured in different ways. We will evaluate our results using confusion matrices. A confusion matrix is a plot showing the true label of the object on one axis and the label predicted by the machine learning algorithm on the other. For the simplest classification problem, a binary classification problem, the confusion matrix would be a $ 2 \times 2 $ matrix with the true positives and true negatives along the diagonal, and the false negatives and false positives on the off-diagonals. Therefore confusion matrices show how well  the algorithm classifies each class. Classes classified correctly would appear on the diagonal, and incorrect classifications would appear on the off-diagonals.


\section{General approach to multiwavelength transient classification}
\label{sec:formal}

Drawing heavily from \cite{Lochner:2016}, we outline a general approach to classifying transients with multimessenger data. The approach is split into two main sections; the first deals with combining data from different sources and the second builds a machine learning classifier that uses light curve data of any wavelength. This creates a general approach useful for combining all sources of information that may be useful to classifying transients, in addition to classification using the light curves themselves.


\subsection{Combining multiple data sources}
\label{sec:multi}

The data used to classify a source need not be information extracted directly from the light curves. It can also be prior or external information about the sources, such as fluxes of the source in different wavelengths or, contextual information, such as position of the object in the sky. Information from alert streams from other observatories can also be added as external information (e.g. the presence of gamma ray emission or a gravitational wave detected by LIGO\footnote{www.ligo.org} in the region of a new radio transient source can be highly discriminating).

There are two methods for incorporating information from other sources:

\begin{itemize}
\item {\bf Probabilistic Approach:} most machine learning classification algorithms are capable of producing a score that can be interpreted as a probability of an object belonging to a particular class. To combine this with external information, such as the presence of a coincident alert at another wavelength, we can calculate the prior probability, $P(\mathcal{C})$, of the object being in a certain class $\mathcal{C}$, given all prior information. This probability, $P(\mathcal{C})$, would then be multiplied by the probability given by the classifier to give a final probability of some object being in class $\mathcal{C}$.

\item{\bf Extra Features:} the second method is to use the information as an extra feature in the machine learning process. For example, if one has a flux measurement at any other wavelength, one could add that flux as a feature. The advantage of this approach is that correlations between the different features are learned automatically by the machine learning algorithm, potentially resulting in improved classification accuracy.

The disadvantage of this latter approach is that \mln{most} machine learning algorithms do not intrinsically deal well with missing data. This could happen if, for instance, MeerKAT detects a transient during a daytime observation when MeerLICHT cannot observe. While feature imputation techniques exist \citep{Quinlan:1993}, a more interpretable approach may be to combine the probabilities where, if data are missing, a default probability based on prior observations (for example, known transient rates) can be used. 
\end{itemize}

The specific setup of the problem will dictate which approach is more appropriate, but formalising this process is a step towards automated multimessenger machine learning pipelines, that can then be fed into downstream analysis including spectroscopic follow-up prioritisation.

\subsection{General approach to transient classification with light curves}
Our method for transient classification follows almost identically the technique outlined in \cite{Lochner:2016}, which was used for classification of supernovae light curves at optical wavelengths. The technique is applicable to any transient (or indeed, \ml{almost} any time series data).\\

\emph{Step 1: Interpolation}\\
\label{sec:step1}

Given some light curve data, $\mathcal{D}$, the first step is to interpolate the data so that it is on a uniform grid. This is done using Gaussian Processes (GPs; \cite{Rasmussen:2005}), since the mean function derived from a GP is an extremely robust interpolator in the presence of noisy data. We use GPs as described in \autoref{sec:gp} for interpolation. \\

\emph{Step 2: Wavelet decomposition}\\
\label{sec:wavelet}

Time series data can be decomposed into a linear combination of basis functions

\begin{equation}
  f(x) = \sum_k a_k \phi_k(x) \; ,
\end{equation}

where $\phi_k(x)$ are orthogonal basis functions and $a_k$ are the respective coefficients.\\

This is a common approach in signal processing and can be a powerful tool for feature extraction, to obtain the set of coefficients used as the features with a machine learning algorithm. One widely used form of this is a Fourier decomposition, where a signal can be decomposed into the component frequencies. However the Fourier decomposition loses all localisation information and is thus mostly applicable to regular, repeating signals.

By contrast, in transient classification, the \ml{object} may be observed at any point in its light curve and the algorithm must determine its class in this setting. Thus, we require a decomposition method that is translation-invariant but still sensitive to the intrinsic shape of the curve. A form of decomposition that is approximately scale and translation-invariant is known as the stationary wavelet transform \citep{Mallat:2009:WTS,Holsch}. Following its successful use in  \cite{Lochner:2016} and \cite{Narayan:2018rxv}, we make use of the \ml{stationary wavelet transform} with the \textit{symlet} family, as implemented in the package \texttt{PyWavelets}\footnote{https://github.com/PyWavelets/pywt}.\\



\emph{Step 3: Dimensionality reduction with PCA}\\

\ml{The stationary wavelet transform produces a large number of redundant features, too many for standard machine learning techniques.} \ksfour{Therefore we need to reduce the number of features while keeping as much information as possible contained in the full feature set.} Principal Component Analysis (PCA) \citep{Pearson:1901} is a dimensionality reduction technique. It is a linear transformation that decorrelates a set of correlated variables by calculating the covariance matrix of the dataset. The eigenvalues and eigenvectors of this matrix are computed.  The total variance of the data can be quantified by calculating the sum total of the eigenvalues. \ksfour{The variance is a proxy for the information contained in the dataset. Thus by maximising the variance of the reduced dataset, the majority of information is conserved.} The eigenvectors with the largest eigenvalues, which describe the majority of the variability in the dataset are stored; smaller eigenvalues are disregarded. \ksn{The number of eigenvectors kept are  decided by the fraction of variability one would want to retain in the dataset; variability is defined as the sum of all eigenvalues. For example, if we want to keep 90\% variability in our dataset, then we would keep the corresponding eigenvectors of the largest eigenvalues (in descending value) until their sum equals 90\% of the sum of all eigenvalues.} The large number of coefficients that we obtain from \texttt{PyWavelets} can be projected onto the stored eigenvectors, producing a new set of eigenvalues.



\section{Application to Radio Transients}
\label{sec:app}
Motivated by the expected increase in new transient detections with modern radio telescopes, we apply our general approach to existing radio transient data. \ksr{Because this data is limited, we use the data interpolation and augmentation technique described in \autoref{sec:gp} as a statistically robust method to simulate light curves for training and testing}. We follow the feature extraction method described in \autoref{sec:formal} and also illustrate the effect of incorporating additional data by including contextual information and optical data.
\begin{figure*}
  \centering
  \includegraphics[width = \textwidth]{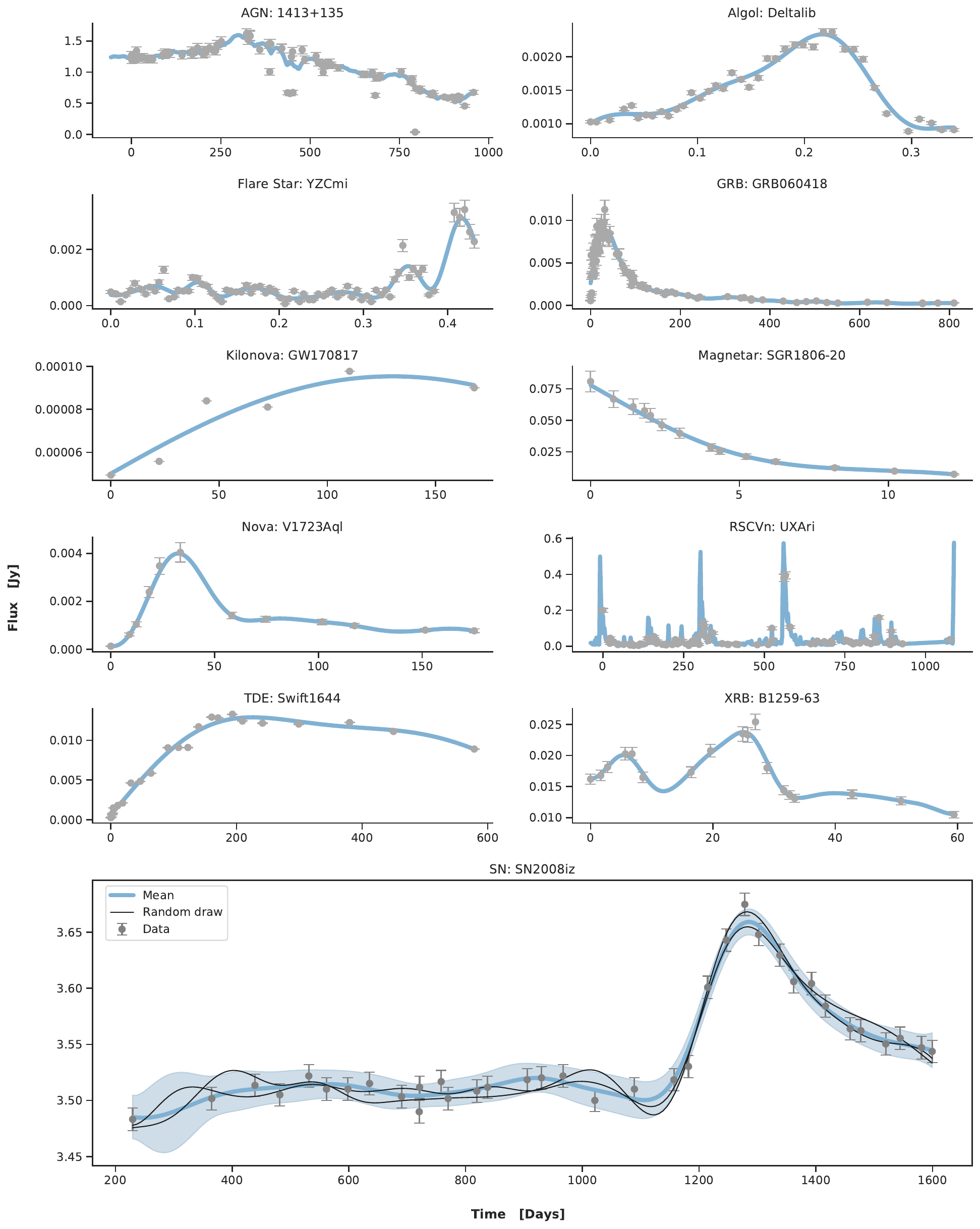}
  \caption{An example light curve for each of the the radio transient types plotted as flux (in Jy) as a function of time (in days). It is clear that the data are taken on extremely different timescales. The original data points are shown in grey and the mean of the Gaussian process is shown in blue. For the SN light curve one standard deviation away from the mean is shown as a blue envelope. The black lines shown in the SN light curve are two random draws from the GP. It can be seen that these lines are different from, but still consistent with the original data.}
  \label{fig:data}
\end{figure*}

\subsection{Radio data}
\label{sec:data}
The radio transient data used were collected by \cite{Pietka:2014} except for the kilonova light curve, which was collected by \cite{Dobie:2018zno}.\\

Most of the data \ksn{are} obtained the literature and the rest is from the Green Bank Interferometer\footnote{https://public.nrao.edu/telescopes/green-bank-interferometer} (GBI). \kss{All the radio observations were done at frequencies between 5GHz and 8GHz.} Data collected from the GBI have a much higher cadence than the data obtained from the literature. The data consists of time series light curve data (radio flux as a function of time). These radio transient light curves can be separated into eleven different classes or types. The total number of light curves in each type and their sources are shown in \autoref{table:data}. An example light curve for each class is shown in  \textcolor{red}{Fig.} \ref{fig:data}.

\begin{table}
  \centering
  \caption{Breakdown of radio transient data into the relevant types and  data sources. GBI refers to data collected from the Green Bank Interferometer. From Lit. refers to data collected from the literature.}
  \begin{tabular}{lccc}
    \hline
    \textbf{Type} & \textbf{From Lit.} & \textbf{GBI} & \textbf{Total} \\
    \hline
    AGN         & 17 & 13 & 30 \\
    Algol       & 1  &  2 & 3  \\
    FS          & 5  &  0 & 5  \\
    GRB         & 4  &  0 & 4  \\
    Kilonova    & 1  &  0 & 1  \\
    Magnetar    & 1  &  0 & 1  \\
    Nova        & 8  &  0 & 8  \\
    RSCVn       & 0  &  2 & 2  \\
    SN          & 13 &  0 & 13 \\
    TDE         & 2  &  0 & 2  \\
    XRB         & 11 &  9 & 20 \\
    \hline
    \textbf{Totals} & \textbf{63} & \textbf{26} & \textbf{89} \\
  \end{tabular}
  \label{table:data}
\end{table}

\begin{figure*}
  \centering
  \includegraphics[width = \textwidth]{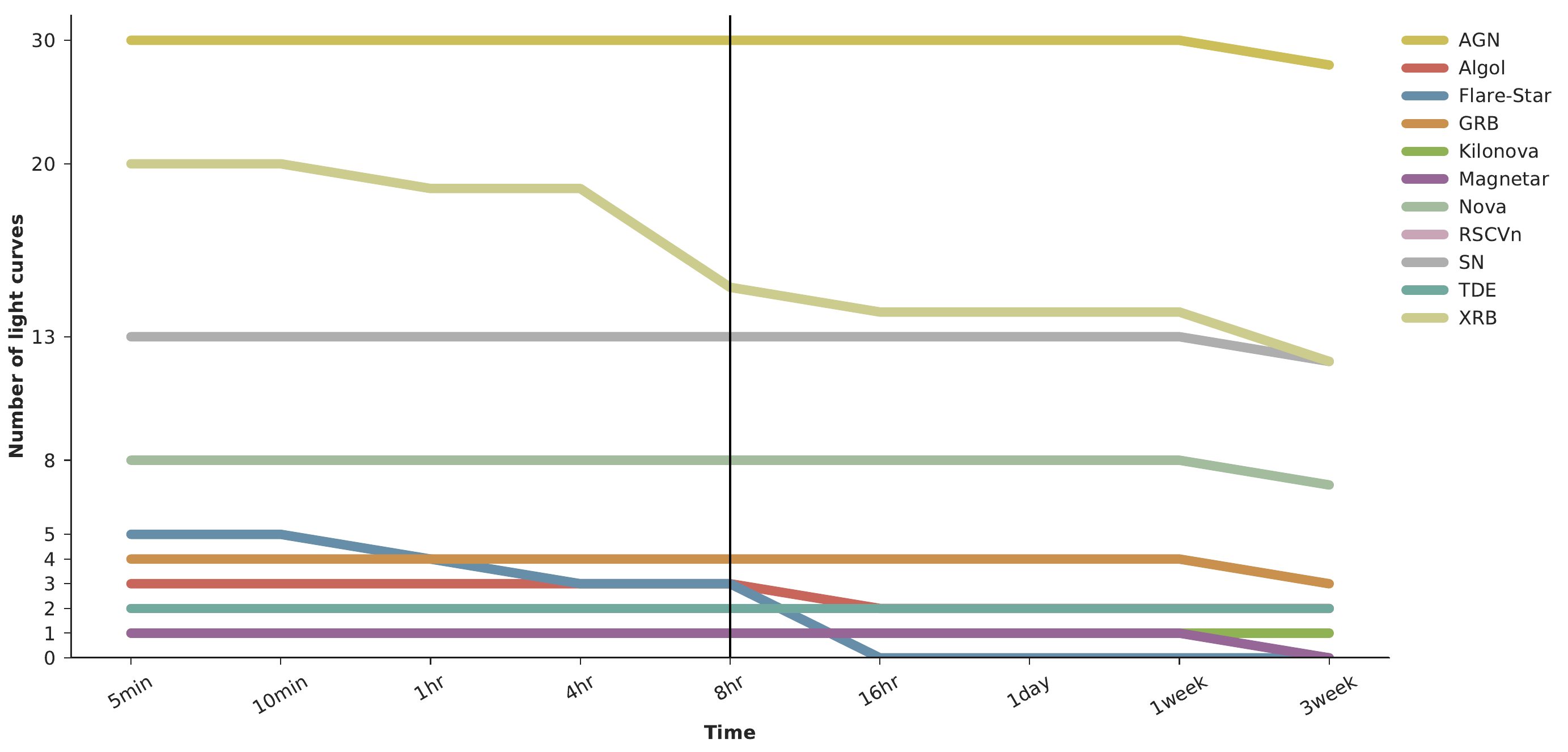}
  \caption{The number of light curves for each class as a function of the length of observation for that that source (in other words, there are $y$ objects with light curves that were observed for at least time $x$). \ksfour{There are clear observational biases here: objects that vary on longer timescales such as AGN or SNe are observed for longer periods and there are greater number of them}. It can be seen that 8 hrs is the shortest observation period for which all the classes have a non-zero number of light curves.}
  \label{fig:numVtime}
\end{figure*}

It is important to note that all the light curves have different lengths. The length of the light curve is correlated with the type of object, due to observational biases. Some objects are observed over years (e.g. AGN) and others are observed over only a few hours (e.g. FS).  \textcolor{red}{Figure}  \ref{fig:numVtime} shows the number of light curves for each class as a function of the total length of observation for that object. Because of this bias, we restrict our study to a timescale of eight hours, which is the longest observation time for which we have measurements for all classes (see  \textcolor{red}{Fig.}  \ref{fig:numVtime}). \ml{Classification on this timescale will also allow relatively prompt follow-up triggers.} The technique is applicable on even shorter timescales, even if there are very few flux measurements, although classification accuracy will likely decrease.

\subsection{Interpolation}
\label{sec:gp}
\ksr{The feature extraction method described in  \autoref{sec:wavelet} requires the input data to be sampled on a regular grid. The data used in this work all have very different cadences hence the light curves need to be interpolated. As mentioned in \autoref{sec:step1}, we use Gaussian processes (GPs) \footnote{Also known as Kriging \citep{1963:Matheron}. } \citep{Rasmussen:2005} to interpolate the light curves due to its robustness in the presence of noisy data.} 

\ksr{GPs provide a stochastic interpolation formalism where the joint distribution of all random variables follow a multivariate normal distribution. Thus every subset of the GP can be characterised fully by its mean and covariance functions. This allows a GP to generate new data that are consistent with the error bars on the observed data. We use \texttt{George} \citep{hodlr} to perform our GP regression.}

\ksr{The characteristic timescales on which light curves vary can be highly class dependant; Supernovae may vary smoothly over months while flare stars may change over an hour. GPs handle such a range of timescales through the covariance function \ksfour{(also known as kernels)}, $k(\Delta t)$, which specifies how quickly the light curves flux values became uncorrelated as the temporal separation between two times, $\Delta t$, increases.}



\ksr{We will use covariance functions that are stationary, that is they depend only on the time difference between data, $\Delta t = |t_i - t_j|$, and a characteristic time scale parameter $\ell$.}



\ksr{In order to get good fits to the wide range of light curves in our data we empirically found that we needed a combination of three covariance functions\ksfour{, or kernels,} to be used for regression, which are defined below.}

\ksr{The first is the Gaussian radial basis function,  given by}
\begin{equation}
  k_\mathrm{rad}(\Delta t) = \exp\left(-\frac{\Delta t^2}{2 \ell^2}\right)\; ,
  \label{eq:rad}
\end{equation}

\ksr{The second kernel we use is the exponential function given by }
\begin{equation}
  k_{\exp}(\Delta t) = \exp\left(-\frac{\Delta t}{\ell}\right)\;  ,
\end{equation}

\ksr{Lastly, we use the exponential sine-squared function,  given by}
\begin{equation}
  k_\mathrm{sine}(\Delta t) = \exp\left( -\Gamma \sin^2\left[ \frac{\pi}{\ell} \Delta t \right] \right)\; ,
\end{equation}

\noindent where $\Gamma$ is a hyperparameter that is allowed to take any value.

\ksr{Given the general temporal variations of the different classes, we found empirically that one of the following three combinations of these kernels provides a good fit to the data:}

\begin{align*}
K_1 &= w_1 k_\mathrm{rad}(\Delta t) \\
K_2 &= w_2 k_\mathrm{rad}(\Delta t) + w_3 k_\mathrm{rad}(\Delta t) k_\mathrm{sine}(\Delta t) \\
K_3 &= w_4 k_\mathrm{rad}(\Delta t) + w_5 k_\mathrm{exp}(\Delta t)
\end{align*}
where the $w_i$ are the weights of each kernel. 

\ksfour{The combinations $K_1$, $K_2$ and $K_3$ were formulated through trial and error. It was found that these combinations fit the dataset well. Due to the current limited dataset, it is likely that there may be new light curves which are not well described by these combinations. More studies would need to be done to test this once a larger dataset has been acquired. We believe that as long as the GP does not overfit, and fits sufficiently well, then the choice of kernel is unlikely to impact the final classification. This is because another feature extraction step is done first, and so is independent of our kernel choice. However, it could have an impact on the data augmentation and this should be studied for a larger dataset.}

\ksr{GP regression on each light curve was performed using each of these combinations, $K_1, K_2, K_3$. We selected the best combination of hyperparameters ($w_i, \ell, \Gamma$) by minimising the negative log likelihood calculated using a dropout test set from the likelihood; a capability built into \texttt{George}}. 

To perform the optimisation of the negative log likelihood we use the Limited memory  Broyden-Fletcher-Goldfarb-Shanno (L-BFGS) optimisation algorithm from the \texttt{SciPy} \footnote{www.scipy.org} package, to \ksn{obtain} the best set of hyperparameters for \ksn{each of the} kernel combinations, \ksr{ $K_1,K_2,K_3$. The $K_i$ with the lowest negative log likelihood was used to construct the GP from which to sample for each light curve. }

\ksr{The advantage of GP regression with a learned covariance kernel, is that it provides an average light curve as well as Gaussian estimates for the uncertainties in the lightcurve at all times, learned from the data of that specific light curve. This is the main advantage over other interpolation methods such as spline, where there is no adaptation to the data.}

\ksr{An example GP for a SN light curve is shown in \textcolor{red}{Fig.} \ref{fig:data}, along with the $1-\sigma$ uncertainty band arising from the covariance kernel specific to this example.}



\subsection{Augmentation}
\label{sec:aug}

\ksr{While the total number of light curves in the our dataset is small, it can be seen from \textcolor{red}{Fig.}  \ref{fig:numVtime} that most of the light curves are longer than one month. Thus one of the ways in which the number of light curves in the dataset can be increased is by splitting the light curves into smaller chunks. GPs fit a function to the original light curves which allows for a statistically rigorous method to sample between any two points $t_1$ and $t_2$ on the light curve provided that $t_2 - t_1 < T$, where $T$ is the total length of the original light curve. This creates new shorter light curves.} \mln{By varying the starting point of the sample, the original light curve can be broken up into smaller chunks.}

Because this method uses GPs, it takes into account the uncertainties of the original dataset which most common augmentation techniques do not. \mln{In addition, we are not restricted to sampling the mean of the GP. We are also able to draw samples from the GP distribution thus creating new light curves that are statistically consistent with the original data.} This method of data augmentation \mln{amounts to} a combination of \cite{trotta:2018} and \cite{2016:Cui}.

\mln{Finally, augmentation is capable of balancing of the classes, artificially enforcing equal numbers of examples in each class. Machine learning algorithms are well known to be biased in the face of highly unbalanced classes and tend to label most objects as the dominant class(es). The use of GPs allows a natural way to construct balanced training sets that do not disfavour the rarer classes.} 

\mln{Appendix \ref{app:aug} includes a rigorous study of the impact of these three aspects of our augmentation technique: the addition of more examples, the balancing of classes and the increase of variance in the training set.}

\subsection{Feature extraction}
\label{sec:feat}

The feature extraction method used is described in  \autoref{sec:wavelet}. First GP regression was performed on the original data set. From this, many example light curves can be generated, each statistically consistent with the original data\ml{, which allows us to generate a realistic synthetic dataset of any size}. \ksr{ First, a reference time, $t_0$, was drawn at random to be somewhere within the original light curve. This was done to simulate the fact that the transient may be detected at any point on its light curve.} We then sampled 100 flux values between $t_0$ and $t_0 + 8\;\text{hrs}$ from the GP. \ml{This approximates an 8 hour radio observation where an image is produced every five minutes.} \ksr{The 8 hour observation time is inline with the expected observing strategy of MeerKAT. } We then used \texttt{PyWavelets} to perform a two-level wavelet decomposition on these 100 points, which returns 400 coefficients. \ksnn{PCA was performed on these, keeping the first 100 principal components. This was done to lessen the computational resources needed to train the classifier.}\\


\subsection{Incorporating external information}
\label{sec:context}

Some classes can have similar light curves but are generally found in different parts of the sky. For example novae tend to occur in the Galactic plane, while SNe are more likely to be extragalactic.

In order to break the degeneracies between these classes we added a feature that characterises where the object is in the sky. Telescopes will always have access to the position in the sky in which it is pointing, hence the drawback of adding features \ml{that may sometimes be missing, as} outlined in \autoref{sec:multi}, will not be an issue.

The RA and Dec coordinates were obtained for all the objects in our dataset using a combination of Simbad\footnote{www.simbad.u-strasbg.fr} and NED\footnote{https://ned.ipac.caltech.edu}. These coordinates were then converted to Galactic coordinates. \ksfour{If we were to copy these coordinates to each of the augmented light curves, the ML algorithm could learn a correlation between the coordinates and the original light curves, which would lead to biased results.}

\ksfour{Therefore, instead we} defined a feature that specified whether or not the object was in the Galactic plane. Any object with a Galactic declination of above $ 10^\circ$ or below  $ -10^\circ$ was considered to be out of the Galactic plane. \ksfour{This new feature (which is either $0$ or $1$) is then appended to the augmented light curve features.}



\section{Results}

\label{sec:results}



\begin{figure*}
  \begin{subfigure}{.5\linewidth}
    \centering
    \includegraphics[width = \columnwidth]{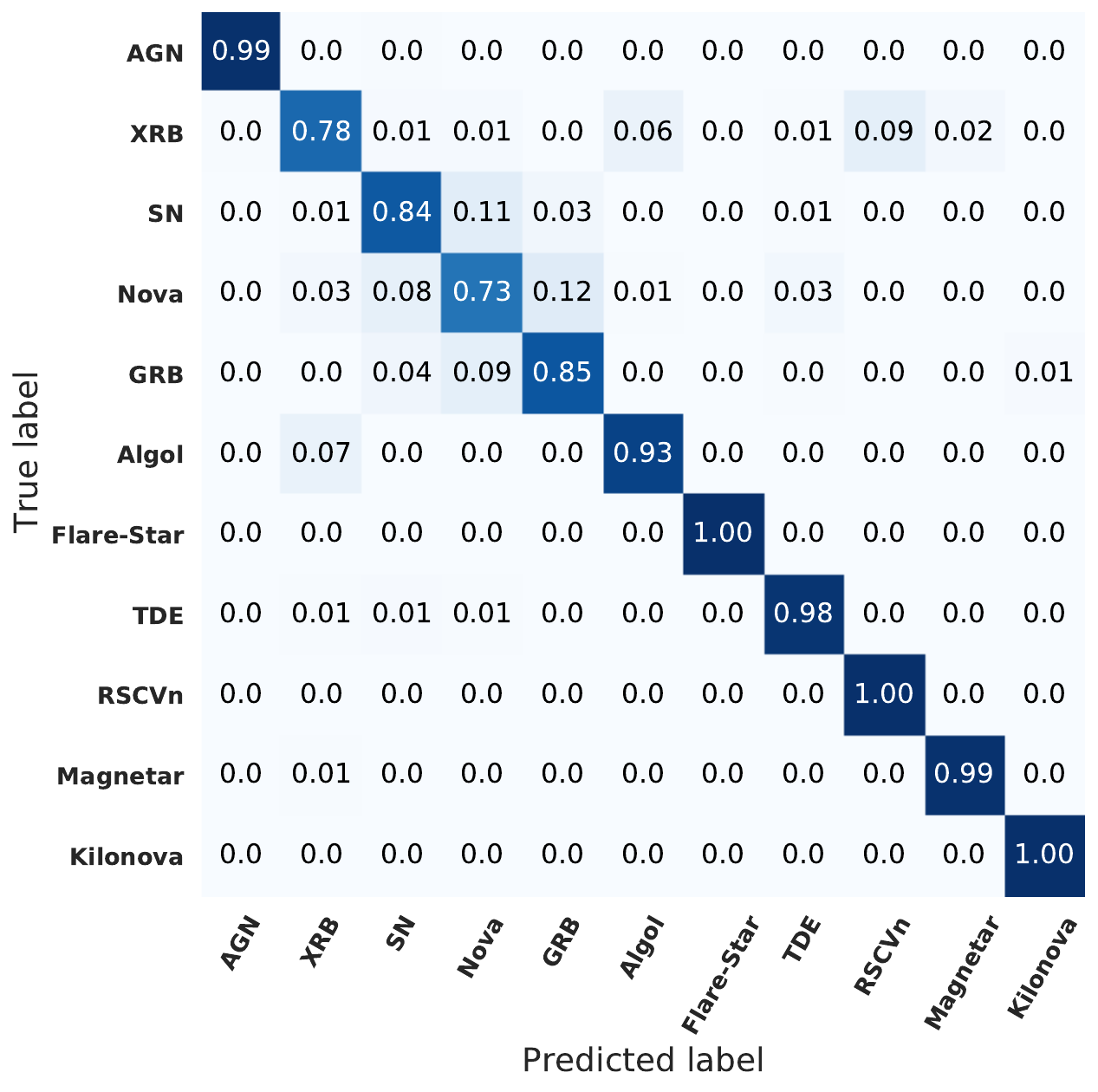}
    \caption{Confusion matrix without contextual information}
  \end{subfigure}%
  \begin{subfigure}{.5\linewidth}
    \centering
    \includegraphics[width = \columnwidth]{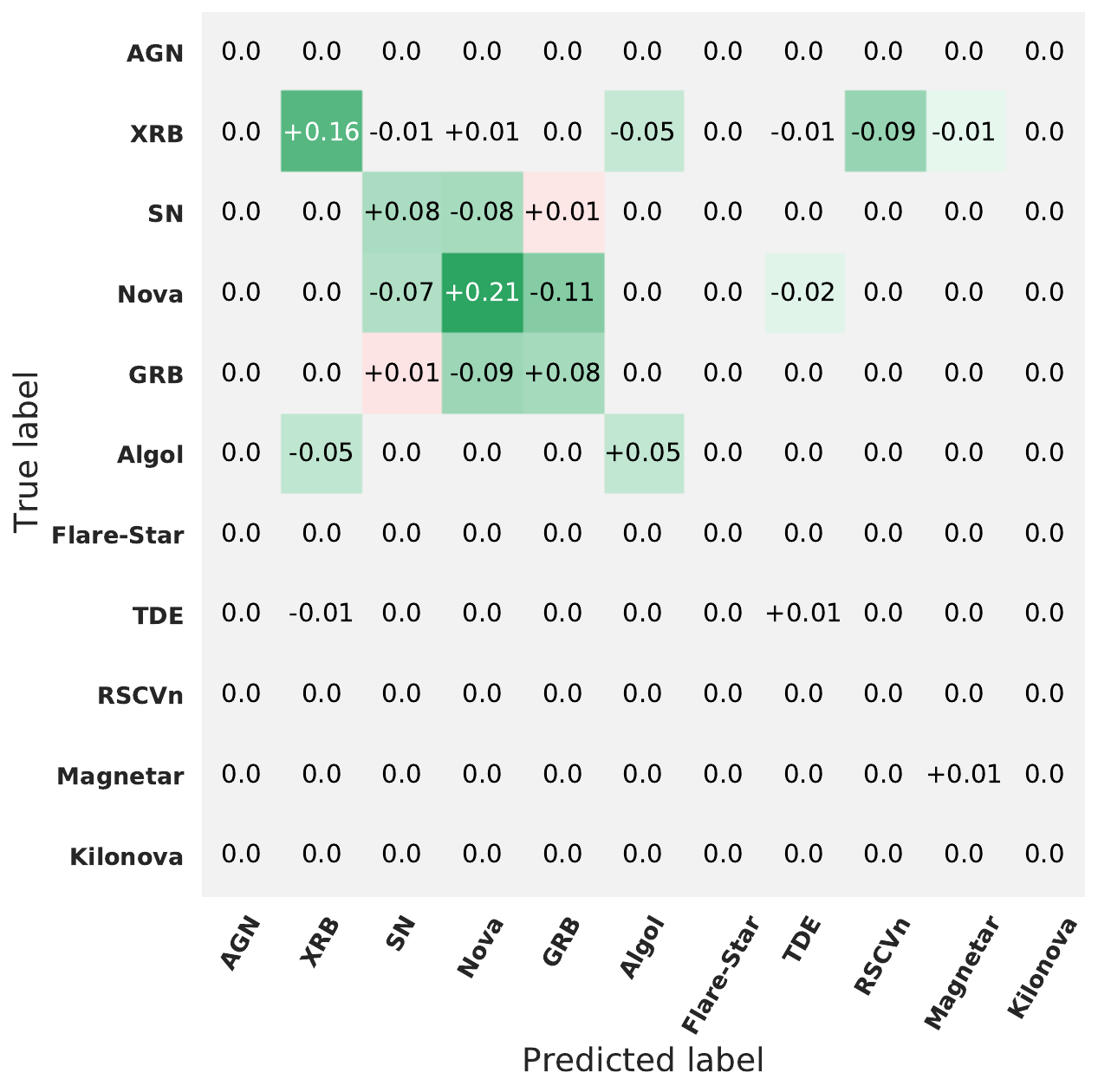}
    \caption{Confusion matrix showing the difference when contextual information is added}
  \end{subfigure}
  \caption{The normalised confusion matrix for wavelet features extracted from eight hours of data. The $y$-axis shows the true label of the object (true class). The $x$-axis shows the label which the algorithm predicts for the object (predicted class). \ksn{ The right panel has two colour schemes. The first corresponds to the diagonals. If the values along the diagonal increase they will show in green; if they decrease they will show in red. The second corresponds to the off-diagonals. If the values along the off-diagonals increase they will show in red; if they decrease they will show in green. From the left panel we see that without any contextual information, the classifier confuses the classes of XRB, SNe, Novae and GRBs. The classifier is greatly improved with the added feature of the object's position on the sky. The four classes are no longer confused as most of the off-diagonals are green. The accuracy for the class of XRBs increased \ksnn{by 16\%, SNe increased by 8\%, Novae increased by 21\% and GRBs by 8\%}} (right panel).}
  \label{fig:cm_w_8}
\end{figure*}

\subsection{The effect of training sets}

\ksr{The radio transient data currently available are unfortunately too small to use as a training set for machine learning. \mln{Machine learning algorithms will tend to generalise poorly from small training sets}. However, by using the method outlined in \autoref{sec:feat} to split up the light curves into smaller chunks and also balance the classes, we are able to build a large enough training set to use for machine learning.  While the simulating of new light curves using GPs does add some variability to the training set, it does not adequately model the real world variability between the different classes. Thus, in order to test the performance of the classifier \mln{in a more realistic setting,} we performed three tests using different training sets. The results of these tests are outlined below.}

\subsubsection{Fully representative training data}
\label{sec:repr}
 \ksr{First, we simulated a representative dataset by training the classifier on samples from all the light curves in our dataset.} GP regression was performed on the original 82 light curves. From these, \ksr{$2,000$} simulated light curves were generated for each of the eleven classes, to create a balanced training set. \ksr{These simulated light curves were eight hours long with 100 evenly spaced data points.} Wavelet feature extraction was then performed on each of the simulated light curves resulting in $400$ wavelet coefficients. After performing PCA on these coefficients, \ml{the 100 most important components were selected and used as features.} \mln{The dataset \ksnn{containing} $2000$ objects \ksnn{per class} was \ksnn{then} randomly split into 75\% training data and 25\% test data.} \ksnn{It is important to note here that $2,000$ light curves were simulated for all classes. This includes the classes with only one object (eg KN) enabling us to do this train-test split across all eleven classes. While this will lead to overfitting to these single light curve classes, it is still important to keep them in the dataset to see which of the other classes the classifier confuses them with.}

\ksr{In order to show the effect of adding contextual information two separate classifiers were trained. One classifier was trained without any contextual feature (i.e whether the source was in or out of the Galactic plane) and the other with contextual information as described in \autoref{sec:context}. \ksnn{During training, we optimised the following three hyperparameters of the random forests algorithm: the number of trees, the splitting criteria and the maximum number of features. A grid search was done for the best hyperparameter values using 3-fold cross validation. The best fitting hyperparameters in this case were found to be 83 trees and a split criterion of entropy with a maximum number of features of 58.}}

The results are shown in \textcolor{red}{Fig.} \ref{fig:cm_w_8}. It can be seen that without any contextual information the classifier confuses the classes of XRB, SNe, Novae and GRBs. \ksr{However after the contextual feature is added, the accuracy for the class of \ksnn{XRBs increases by 16\%, SNe increases by 8\%, Novae increases by 21\% and GRBs by 8\%. Thus the overall average increase in accuracy is $\approx 7\%.$}} \ksfour{It can also been seen that while the overall accuracy increases, some classes such as GRBs and SNe get more confused. This is because these classes were previously confused with novae. With the introduction of the external information feature, they clearly separate away from novae but are now more likely to be confused with each other instead. However, the overall confusion between classes is reduced.} We expect the confusion matrix with contextual information to characterise the performance of the classifier in practice; \ml{thus this contextual feature was used in the rest of this work}. 

\ksnn{ It can also be seen that classes with very few original light curves (eg KN, RSCVn) have perfect accuracies. This is due to over fitting. The training and test sets for these classes were simulated from a small number of original light curves (between one and three originals). Therefore these training and test sets are unlikely to have much variation hence making it easy for the classifier to over fit to them.}

\subsubsection{Non-representative training set}
\label{sec:drop20}
\ml{Because our original data set is limited, we expect the results outlined in \autoref{sec:repr} to represent an idealised case. In practice, we anticipate that any test set would consist of some objects not present in our training set.} To construct a more realistic test of the classifier performance with the data currently available, instead of training the classifier on samples from all the original light curves, we trained it on a subset of light curves. As can be seen from \textcolor{red}{Fig.} \ref{fig:numVtime}, only five classes have greater than four light curves. In order to ensure the training subset contained all classes, we removed light curves from only these five classes. We still included the classes with only one object in both the training and test sets, because these objects can cause confusion between classes which would be artificially removed if they were excluded. 

We removed 25\% of the light curves in these five classes at random, GP regression was performed on the remaining 75\% of the original light curves. From these, \ksr{$1500$} simulated light curves were generated for each of the 11 classes. \ksr{As before, these simulated light curves were eight hours long with 100 evenly spaced data points.} Wavelet feature extraction was then performed on each of the simulated light curves followed by PCA\ksr{, resulting in \ksnn{100} features as before.} This was used as the training set. \mln{A further \ksr{$500$} simulated light curves were drawn from the remaining 25\% of the light curves. GP regression, wavelet extraction and the PCA components from the training set were used to project the high dimensional feature set down to \ksnn{100} features as before.} This was used as the test set. \ksr{The random forest was trained and the hyperparameters optimised as described in \autoref{sec:repr}.}

\begin{figure}
\centering
\includegraphics[width = \columnwidth]{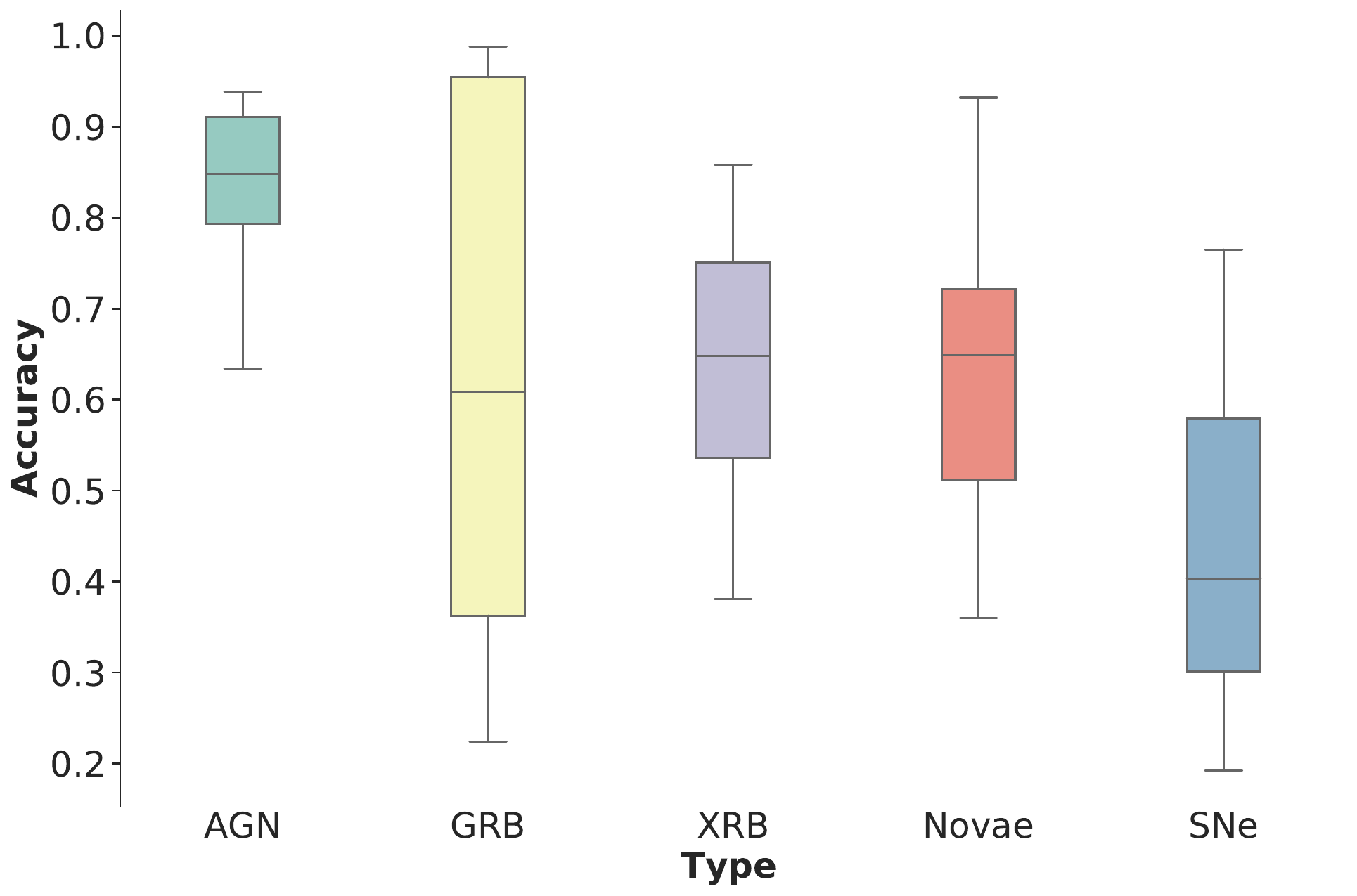}
\caption{\ksn{Summary of results from non-representative training set are shown in these box and whisker plots. The bar in each box shows the accuracy for each class averaged over the 20 runs (on each run a classifier was trained on 75\% of the total dataset and tested on the remaining 25\%, this process was repeated 20 times, each time randomising the training and testing sets). The coloured boxes show the interquartile range. The grey bars (whiskers) show the minimum and maximum accuracies in the 20 runs. It can be seen that the classifier performs poorly for SNe as it has the lowest average accuracy of $\approx 40\%$.}}
\label{fig:min_max}
\end{figure}

\begin{figure}
\centering
\includegraphics[width = \columnwidth]{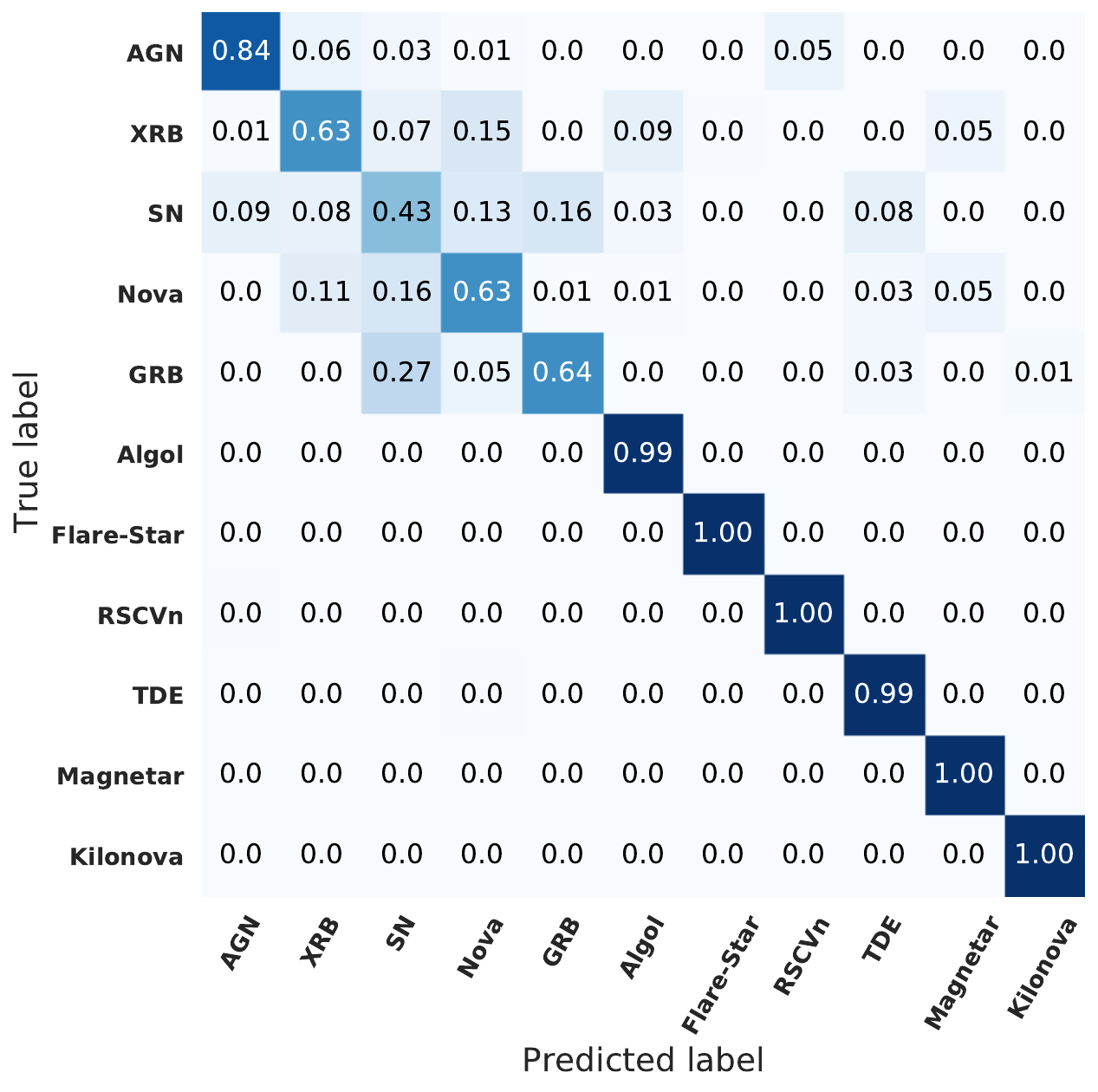}
\caption{Confusion matrix showing the results of the classifier with a non-representative training set averaged over 20 runs. The $y$-axis shows the true label of the object (true class). The $x$-axis shows the label which the algorithm predicts for the object (predicted class). \mln{The classifier performs well on well-observed classes such as AGN but poorly on SNe which have a small training set with large variability. The classifier performs artificially well on classes with single examples. We expect the performance of the classifier on current real world data to be similar to that shown here.}}
\label{fig:real}
\end{figure}

This was repeated 20 times, each time removing 25\% of the light curves at random. The results of this is shown in \textcolor{red}{Figs.} \ref{fig:min_max} and \ref{fig:real}. \ks{It can be seen that the performance of our classifier has decreased \ksnn{by $\approx $ 19\%}, which is to be expected given the non-representativeness of the training set.} \ml{In particular, well-represented classes like AGN still perform well but classes such as SN, where the training set is highly diverse with many dissimilar objects, are poorly classified.} \ksnn{The overall accuracy averaged over the 20 runs was found to be 78\%. We expect the performance of the classifier on real world data, which has similar properties to our dataset, to be similar to the performance on these held out datasets.}

\subsubsection{Single curve testing}

To further demonstrate the effect of a non-representative training set, we tested the classifier's performance on light curves not represented in a training set. We focused on the five classes  (AGN, SNe, XRBs, Novae and GRBs) that contain four or more light curves.

We trained the classifier on samples from all \ml{of the original} light curves, except for one. GP regression was \ml{thus} performed on 82 of the original light curves \ml{and} from these simulated light curves were drawn, \mln{that were eight hours long with 100 evenly spaced data points.} Wavelet feature extraction was then performed on each of the simulated light curves followed by PCA. \ksnn{Unlike in the previous two tests, we set the number of features to 20 (from the PCA, which corresponds to  retaining $99\%$ of the variability in the dataset). One could also perform a hyperparameter optimisation over the max feature parameter, as we did in the previous two sections, to find the best number of features to use. While this parameter should ideally be optimised, we did not have the computational resources to optimise all possible hyperparameters and set them instead to sensible defaults. We ran a small scale test optimising the maximum number of features and found it increases accuracy by a negligible amount.}

We then \ksn{tested} the classifier on simulated light curves from the one \ksnn{original light curve} that was left out of the training set in the training process. This process was then repeated, each time omitting a different light curve in the training set, until every light curve in the dataset had been removed during the training of a classifier at least once. \mln{Because we make use of data augmentation, each light curve actually represents a large sample of smaller, augmented light curve. We can thus evaluate the anticipated accuracy with which this object would be classified.} 

\begin{figure*}
  \centering
  \includegraphics[width = \textwidth]{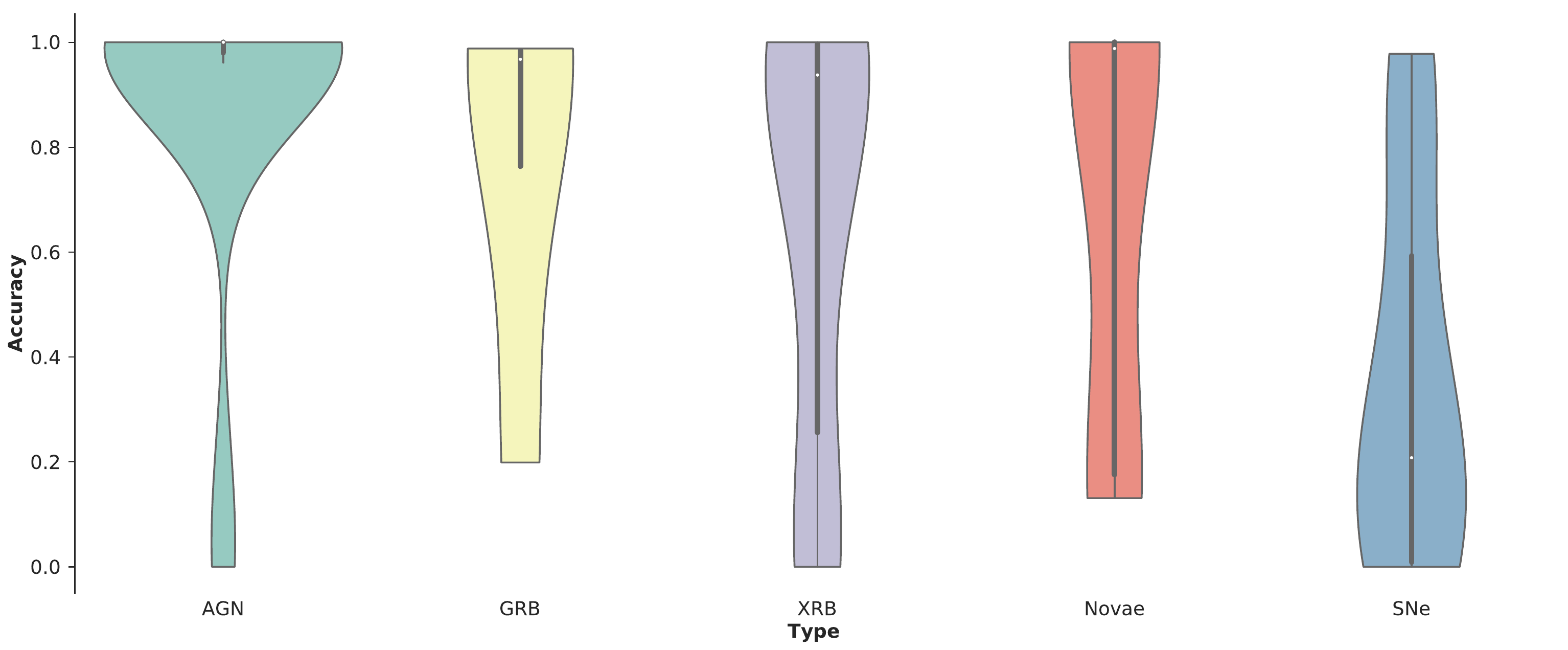}
  \caption{The single curve testing results for the five main classes are summarised in these violin plots. Each violin plot represents a different class as shown. The coloured outline shows the smoothed kernel count distribution of the accuracies. The thick central black line represents the interquartile range. The thin central black line shows the 95\% interval. From this is can be seen that the classifier performs well with AGN as most of the accuracies are above 80\%. It can also be seen that the classifier performs poorly with SNe as most of the accuracies are below 50\%. \mln{There is notable spread in all the classes, which is a symptom of the small training set and will improve with additional data.}}
  \label{fig:violin}
\end{figure*}

The accuracy for each of the dropped out curves is shown in \autoref{tab:drop}. The results for these five classes are summarised in the violin plots shown in \textcolor{red}{Fig.} \ref{fig:violin}. The coloured outline shows the smoothed kernel count distribution of the accuracies. The thick central black line represents the interquartile range. The thin central black line shows the 95\% interval. From this is can be seen that the classifier performs well for AGN as most of the accuracies are above 80\%. It can also be seen that the classifier performs poorly for SNe as most of the accuracies are below 50\%. \ksfour{\textcolor{red}{Figures} \ref{fig:min_max} and \ref{fig:violin} both show that SNe performs worse than all the other classes. This suggests that SNe exhibit greater intrinsic variability than other classes in our dataset. This can be easily confirmed by visual inspection of the data: we found that most SNe light curves were visually fairly distinct from each other. With a small dataset that suffers from observational bias, it is difficult to conclude whether or not this variability is intrinsic to the SNe population. We can however suggest that this would be the most important class to increase the size of the training set for, to improve this and any future classifiers.}

\kss{From \autoref{tab:drop} and \textcolor{red}{Fig.} \ref{fig:violin} we can see that all the classes except for AGNs have a large variance, this means that the classifier has not adequately learnt the underlying properties of these classes. This is due to the fact that these classes do not have enough examples. It is important to note however, that even with a few examples the classifier can perform well, for example GRBs perform better and have a lower variance than SNe despite SNe having more examples. This is because the four GRB examples in this dataset are relatively similar to each other while the 13 SNe examples are not hence the classifier can learn the underlying properties of the GRBs and not the SNe. Therefore we can expect the classifier to perform well on new objects if they are similar to those in the training data. However given how small this dataset is, it is unlikely that new objects found would be similar thus we do not expect this classifier to generalise well until there is a sufficiently large training set. }

\begin{figure*}
\begin{subfigure}{.5\linewidth}
\centering
\includegraphics[width = \columnwidth]{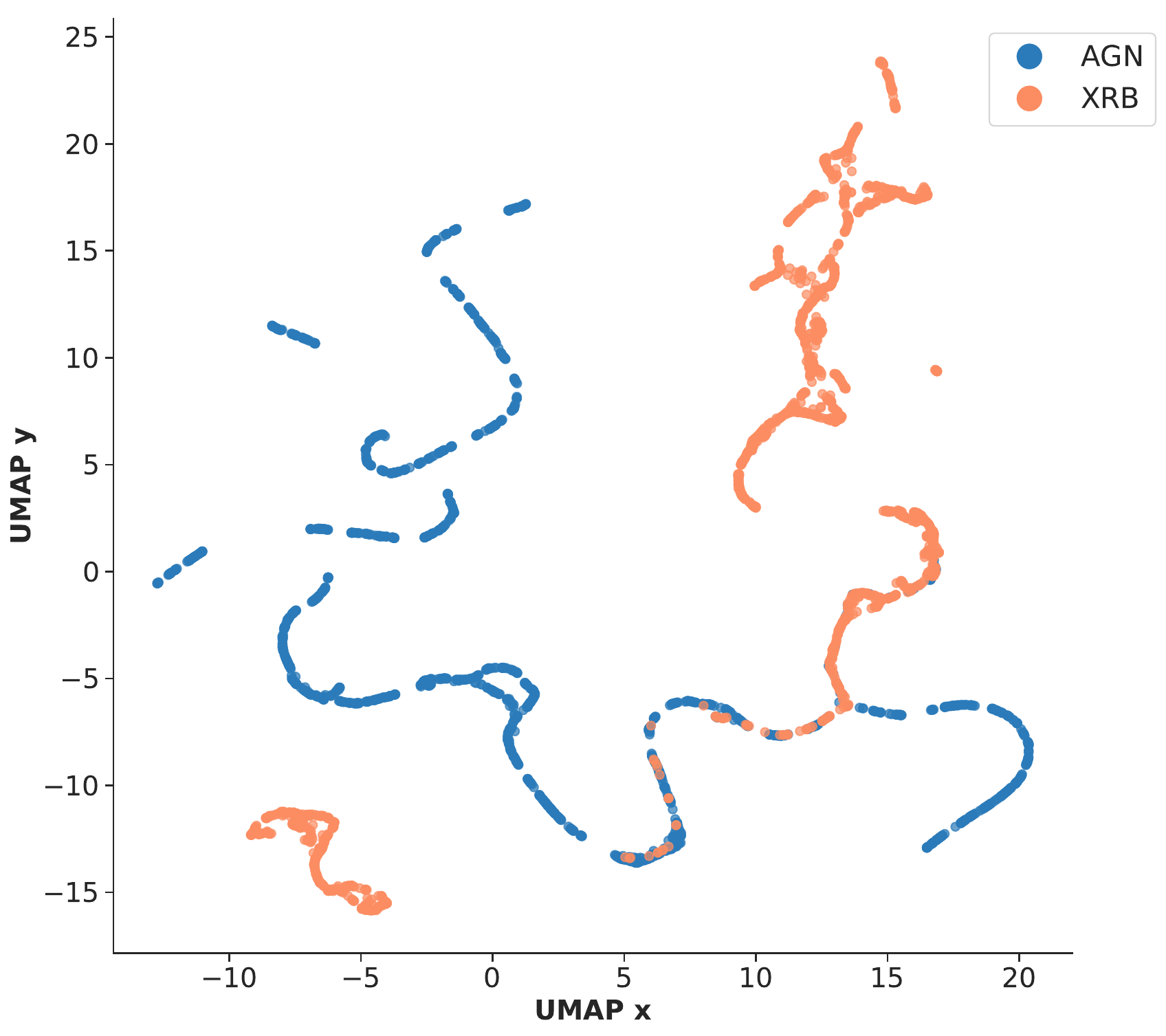}
\caption{AGN-XRB UMAP}
\end{subfigure}%
\begin{subfigure}{.5\linewidth}
\centering
\includegraphics[width = \columnwidth]{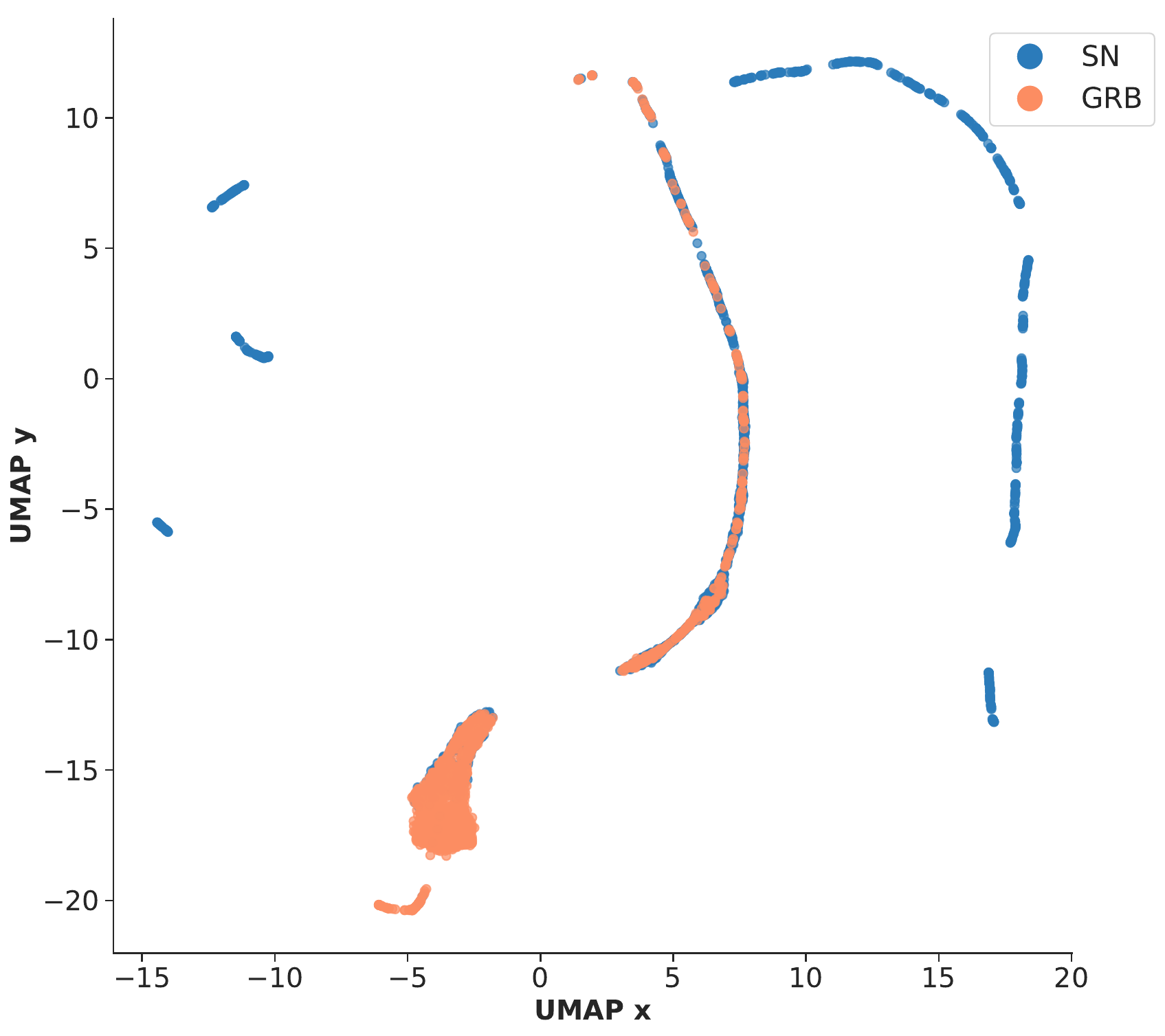}
\caption{SN-GRB UMAP}
\end{subfigure}\\[1ex]
\caption{\kss{ UMAP plots for the four classes of radio transients are shown. UMAP  plots show the high dimensional feature space embedded in two dimensions, hence the x and y axes are arbitrary.  The left panel shows the UMAP plot for two classes for which the classifier performs well. The blue points represent feature vectors of AGN objects, orange points represent the feature vectors of XRB objects. The right panel shows the UMAP plot for two classes for which the classifier performs poorly. The blue points represent feature vectors of SN objects, orange points represent the feature vectors of GRB objects. It can be seen that features from classes that the classifier does well with are relatively separated in feature space (left panel) where as features for classes which the classifier confuses overlap in feature space (right panel). }}
\label{fig:umap}
\end{figure*}

As described in \autoref{sec:vis}, the UMAP plots show the distribution of features in high order feature space in two dimensions. It can be seen from \textcolor{red}{Fig.} \ref{fig:umap}  that features from classes where the classifier performs well are relatively separated in feature space. However \ml{data} for classes which the classifier confuses overlap in feature space hence the classifier is unable to tell the difference between these objects.

\subsection{Adding optical data}

\begin{figure*}
  \centering
  \includegraphics[width = \textwidth]{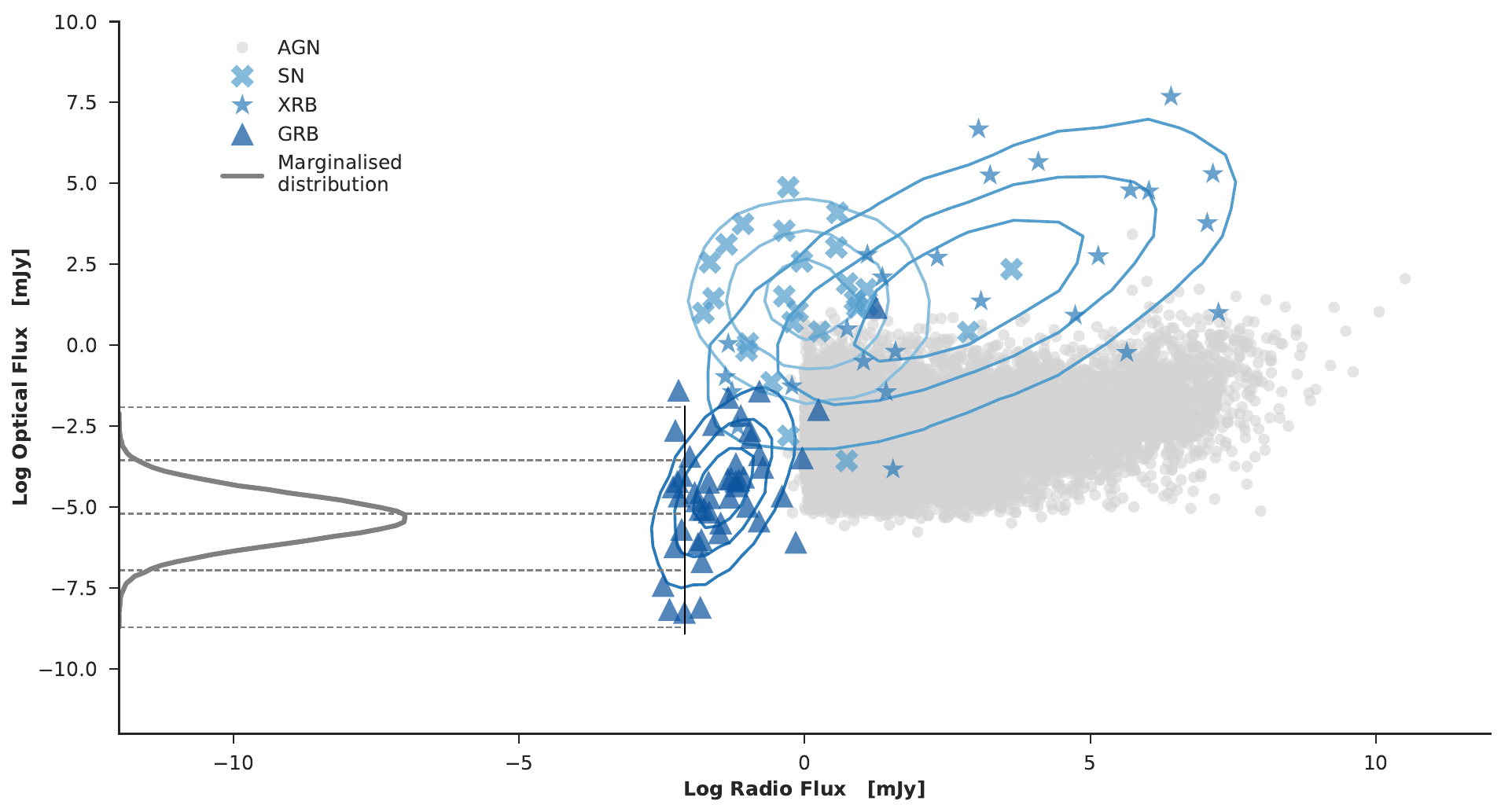}
  \caption{Optical - radio flux distributions for the classes of AGN, XRB, GRB and SNe. Two dimensional Gaussian fits to the distributions of three of the classes are shown with contours. These fit the data for the three classes reasonably well, however it can be seen that the distribution for AGNs is highly non-Gaussian. An example peak GRB radio flux measurement is shown as a black vertical line. The GRB radio-optical flux distribution marginalised over this peak flux is shown in dark grey. Optical fluxes were sampled from these marginals.}
  \label{fig:dist}
\end{figure*}

Simultaneous optical observations do not exist for all the classes in our dataset. To show how the addition of multiwavelength data could help in classification, an optical flux measurement had to be simulated. Optical data for four classes, namely: AGN, XRB, SNe and GRB, were collected from \cite{Stewart:2018ket}. The data consisted of single flux measurements in both optical and radio wavelengths for each source. The dataset had total of  $11,882$ measurements of which $11,782$ were AGN. \textcolor{red}{Figure} \ref{fig:dist} shows the radio-optical flux distributions for each of the four classes. A two dimensional Gaussian was used to model the distribution all of the classes except for AGN which can be clearly seen to be highly non-Gaussian. The Gaussian fits are shown as contours. These Gaussian distributions were used to sample new optical fluxes for the three classes, new optical fluxes were sampled directly from the distribution for AGNs as there \ksn{are $\approx 11\;000$ data points}. 

\ksn{Using the class of GRB as an example, the process for simulating simultaneous optical and radio observations is is follows. First,} a GRB radio light curve was simulated as described previously; the peak flux of this light curve was found. The radio-optical flux distribution was marginalised over, given the peak radio flux, as shown in black in \textcolor{red}{Fig.} \ref{fig:dist}. An optical flux was then drawn from this marginal distribution. Finally, this optical flux was added as an extra feature in the machine learning process. \ks{For the class of AGN, a peak radio flux was found as before, points in the radio-optical flux distribution was then binned, centered on the peak radio flux with a bin width of $0.1$ mJy. These binned points, now marginalised over the peak radio flux follow a Gaussian distribution similar to that of the other classes. An optical flux was then drawn from this marginal distribution and added an extra feature in the machine learning process.}


\begin{figure*}
  \begin{subfigure}{.5\linewidth}
    \centering
    \includegraphics[width = \columnwidth]{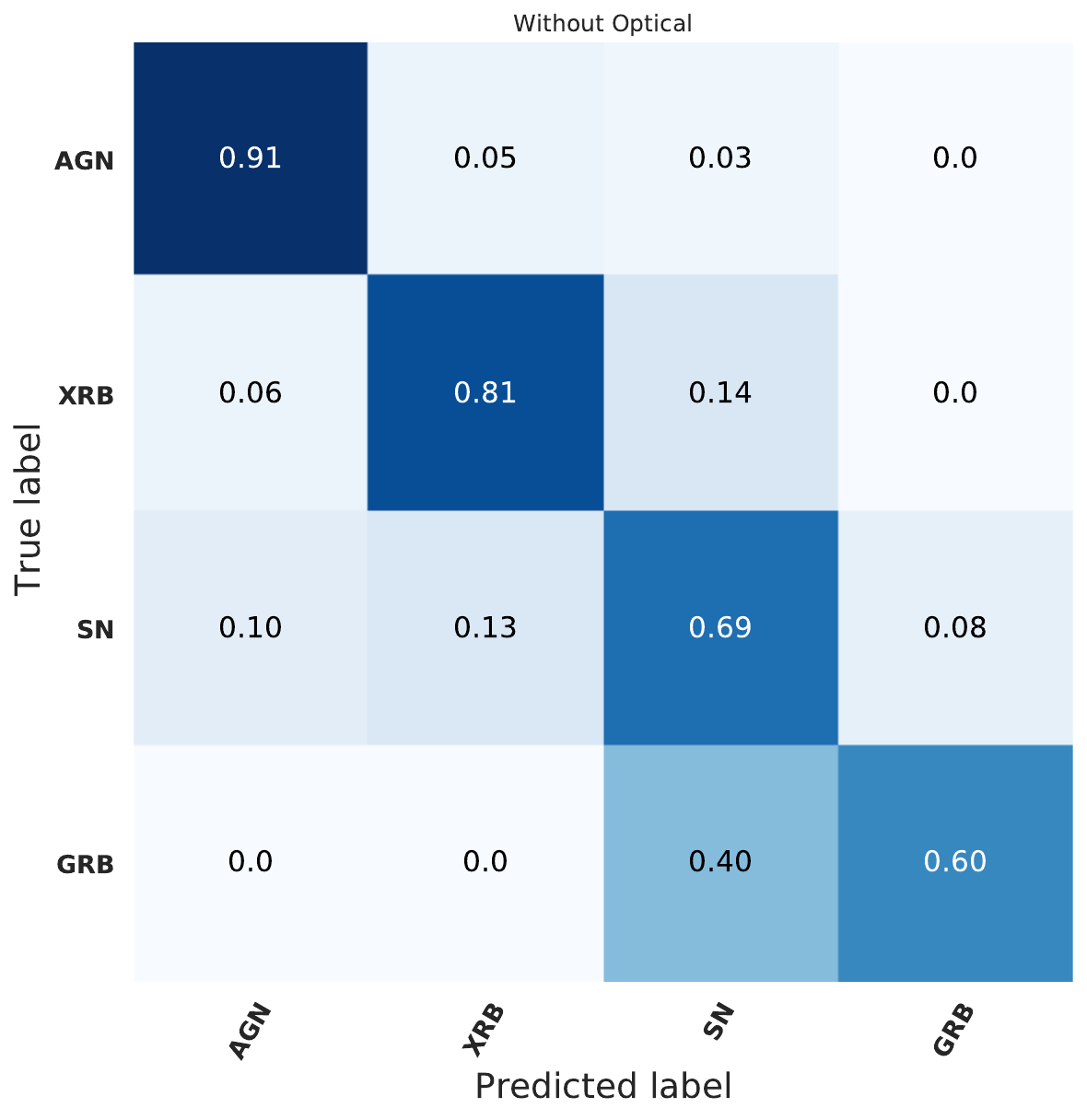}
    \caption{Confusion matrix without optical feature}
  \end{subfigure}%
  \begin{subfigure}{.5\linewidth}
    \centering
    \includegraphics[width = \columnwidth]{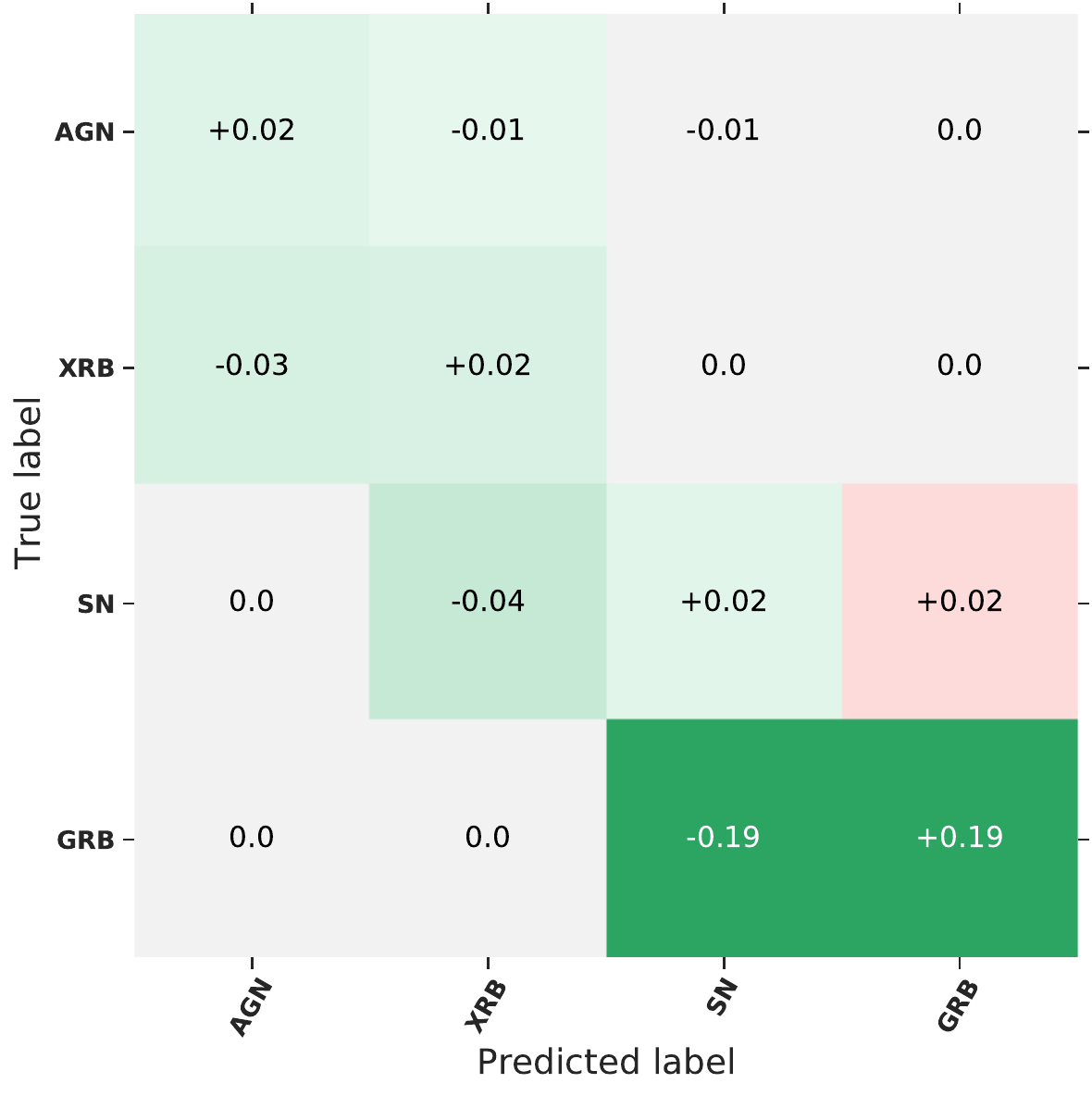}
    \caption{Confusion matrix showing the difference when optical feature is added}
  \end{subfigure}
  \caption{The normalised confusion matrix for wavelet features extracted from 8 hrs of data. The $y$-axis shows the true label of the object (true class). The $x$-axis shows the label which the algorithm predicts for the object (predicted class). \ksn{The right panel has two colour schemes. The first corresponds to the diagonals. If the values along the diagonal increase they will show in green; if they decrease they will show in red. The second corresponds to the off-diagonals. If the values along the off-diagonals increase they will show in red; if they decrease they will show in green. From the left panel it can be seen that without the optical feature the classes of GRBs and SNe are confused. With the added optical feature the performance of the classifier is greatly improved as \mln{most of} the off-diagonals are green. From the diagonals it can be seen that most significant improvement in performance \ksnn{is the class of GRBs which increase in accuracy by 19\%}} (right panel).}
  \label{fig:optical}
\end{figure*}

The method outlined in \autoref{sec:drop20} was repeated using the four classes mentioned above. The average confusion matrices with and without an optical feature are shown in \textcolor{red}{Fig.}  \ref{fig:optical}.
It can be seen that the addition of this optical feature improves the performance of the classifier.  The average accuracy increases by 5\% with the biggest gain seen in the worst performing class, the GRBs, where the number of GRBs misclassified as SNe has dropped by 19\%. \ksfive{We note that because our dataset of GRBs is drawn from just four light curves, this may exaggerate the improvement in performance for this particular class. Thus more data would be needed to confirm the magnitude of this improvement. Gains in other classes are more modest, in the 2\% range. It's interesting to note that with the added optical data, SNe are now slightly (2\%) more likely to be confused as GRBs. Thus there is an asymmetry in the off-diagonal terms of the confusion matrix for these two classes. This could be due to the inherent variability of the two classes in our dataset, which has a stronger than usual impact due to the small number of light curves in our original (unaugmented) data set. Because there are more SNe in the original data than GRBs, the variability of that class is better covered (and thus likely to be somewhat more realistic) meaning it is more difficult to improve classifier performance with the addition of a simple optical feature. Overall however, these results indicate that including optical data is a promising way to improve classification performance to some degree across all classes.} This improvement in performance demonstrates the value of simultaneous optical observations, such as from MeerLICHT in classifying radio transients.



\section{Conclusions}
\label{sec:conc}
We have presented a general \ksr{approach} for multiwavelength transient classification with machine learning. \mln{We have outlined multiple ways to include data from different} telescopes, as well as additional information such as source location. 

Extending \cite{Lochner:2016}, we developed a machine learning pipeline for classifying radio transients. We have illustrated how to include contextual information (such as whether or not the source is in the Galactic plane) and simultaneous observations from an optical telescope. \ksr{      This is, to our knowledge, the first case of applying a machine learning formalism for radio transient classification with real data. \mln{While our sample size is unfortunately only 82 light curves, we have rigorously investigated the effect of this and} have showcased a new data augmentation technique (based on \cite{trotta:2018, 2016:Cui}, specifically to address, in a statistically rigorous way, the otherwise perennial issue of a small sample size.} 

We tested our pipeline using existing radio transient light curves gathered from literature. Because these light curves were few in number, we artificially augmented the dataset using Gaussian processes. \mln{Using Gaussian processes as an interpolation and augmentation technique allowed us to split up long light curves into smaller 8-hour examples for training, balance the classes and introduce a small amount of additional variation statistically consistent with the original data (see Appendix \ref{app:aug} for more details).} Features were extracted from this augmented dataset using a wavelet decomposition, followed by dimensionality reduction. We found that the wavelet features have much higher performance than a simpler set of features based on the change in flux of the light curve over different time periods \mln{(see Appendix \ref{sec:simple_res} for further details)}. 

\ksnn{We have shown that the classification accuracy increases for some classes when contextual information is included as an additional feature.  The improvement is particularly noticeable for novae, objects typically found within the galactic plane, which are otherwise confused with GRBs and SNe.  }

If the \ksn{training data are representative of the test data}, the performance of the algorithm is excellent, achieving an average accuracy of \ksnn{97\%}. However, because our training set is so small, it is highly likely that sources observed by new telescopes such as MeerKAT would not be similar to anything in the training set.

If we remove 25\% of the objects from the training set and then test on those \ksn{removed} light curves, we achieve an average accuracy of only \ksnn{78\%}. \ksnn{We expect this to be the more realistic performance of the classifier while training sets are small.} From the investigation of the effect of dropping out single individual light curves we find that, while most of the light curves are still classified well, several light curves in the dataset are \ksn{poorly} classified. This is because they are dissimilar to anything in the training set. This effect is most pronounced for supernovae, which are generally easily confused with other classes such as GRBs or novae. \mln{This implies that obtaining training data for such difficult to classify objects should be prioritised as part of modern radio transient surveys.}

Finally, we illustrate the effect of including optical data. Unfortunately simultaneous optical and radio transient data are scarce, so we simulate optical fluxes for the radio light curves, drawing on existing distributions. \ksnn{Using the more realistic ``dropped out'' training set, we find that the addition of the optical feature dramatically reduces the confusion between SNe and GRBs.}

\kss{Our extensive analysis shows that while the technique performs well and small training sets can be utilised, it is still limited by the availability of a representative training set. This is difficult to obtain due to few examples of radio transients as well as biases in selection effects, telescope sensitivity etc. \ksfive{In general, it can be difficult to know when one has enough training examples of a given class to provide reliable classification. Classes exhibiting more intra-class variability generally require more training examples. Our approach outlined in Sections 5.1.2 and 5.1.3 can assist in determining this since the classifier performance will not vary greatly as subsets of data are removed once there are sufficient training examples for that class. Additionally, one should monitor the performance of the classifier as object classes are spectroscopically confirmed. While small training sets are a challenge in any machine learning setting, we do expect that the algorithm will scale well and achieve good performance with real data, with classification performance and reliability increasing as more training data are added.} For use on a modern telescope such as MeerKAT, we recommend the classifier be used in online mode where new training examples are added continuously, thus improving performance of the algorithm.}

These results indicate that by including multiwavelength information and making use of a sophisticated machine learning approach, we can expect accurate classification of radio transients with even a small training set created with early MeerKAT and MeerLICHT data.


\section*{Acknowledgements}
We acknowledge support from SKA Africa and the National Research Foundation (NRF) towards this  research. Opinions expressed and conclusions
arrived at, are those of the authors and are not  necessarily to be attributed to the NRF. We acknowledge support from STFC for UK participation in LSST through grant ST/N00258X/1 and travel support provided by STFC for UK participation in LSST through grants ST/L00660X/1 and ST/M00015X/1. HVP was supported by the European Research Council under the European Community's Seventh Framework Programme (FP7/2007-2013)/ERC grant agreement no 306478-CosmicDawn. PAW acknowledges support from the University of Cape Town. OL acknowledge STFC Grant ST/R000476/1.



%
%
\appendix

\section{Tables}

\begin{table*}
  \centering
  \caption{The results of the individual dropout test. Samples from one light curve was removed during training. These samples were then used to test the  classifier. The accuracy for the samples of each of the light curves for the five main classes are shown. From left to right the classes are: AGN, XRB, SN, Nova and GRB. From these tables it can be seen that the classifier performs well for AGNs, XRBs and GRBs but poorly for SNe and Novae. This shows that the intrinsic variabilities of these two classes are not captured by our dataset. }
  \setlength\tabcolsep{4pt}
  \begin{tabular}{ccccc}
    \begin{tabular}[t]{l|c}
      \hline
      \textbf{Name} & \textbf{Acc} \\
      \hline
      NGC7213 & 0.0 \\
      0850-121 & 0.0 \\
      0954+658 & 0.206 \\
      0224+671 & 0.784 \\
      2005+403 & 0.824 \\
      NRAO530 & 0.931 \\
      2223-052 & 0.951 \\
      1413+135 & 0.961 \\
      0528+134p & 0.98 \\
      1622-297 & 1.0 \\
      CTA102 & 1.0 \\
      0336-019 & 1.0 \\
      3C345 & 1.0 \\
      B0605-085 & 1.0 \\
      3C454.3 & 1.0 \\
      1328+254 & 1.0 \\
      3C120 & 1.0 \\
      3C273 & 1.0 \\
      0458-020 & 1.0 \\
      3C279 & 1.0 \\
      1237+049 & 1.0 \\
      0851+202 & 1.0 \\
      2200+420 & 1.0 \\
      0528+134 & 1.0 \\
      PKS2004-447& 1.0 \\
      1803+784 & 1.0 \\
      0954+65 & 1.0 \\
      1749+096 & 1.0 \\
      NGC4278. & 1.0 \\
      AO0235+164 & 1.0 \\
    \end{tabular}
    &
    \begin{tabular}[t]{l|c}
      \hline
      \textbf{Name} & \textbf{Acc} \\
      \hline
      B1259-63 & 0.0 \\
      MAXIJ1836 & 0.007 \\
      ScoX-1 & 0.014 \\
      CygX-2 & 0.148 \\
      XTEJ1550& 0.581 \\
      aqlX1 & 0.707 \\
      GROJ1655 & 0.927 \\
      1909+048 & 0.948 \\
      SS433 & 0.954 \\
      0236+610 & 0.981 \\
      1915+105 & 0.986 \\
      GX17+2 & 1.0 \\
      CygX-1 & 1.0 \\
      CICam & 1.0 \\
      CirX-1 & 1.0 \\
    \end{tabular}
    &
    \begin{tabular}[t]{l|c}
      \hline
      \textbf{Name} & \textbf{Acc} \\
      \hline
      SN1993J & 0.0 \\
      SN2008iz & 0.0 \\
      SN1980K & 0.0 \\
      SN1988z & 0.009 \\
      SN2003L & 0.147 \\
      SN1998bw & 0.195 \\
      SN2003bg & 0.208 \\
      SN2011dh & 0.307 \\
      SN2004dk & 0.593 \\
      SN1994I& 0.593 \\
      SN2008ax & 0.861 \\
      SN2004cc.& 0.931 \\
      SN2004gq& 0.978 \\
    \end{tabular}
    &
    \begin{tabular}[t]{l|c}
      \hline
      \textbf{Name} & \textbf{Acc} \\
      \hline
      V1974Cyg& 0.131 \\
      RSOph & 0.152 \\
      V1500Cyg& 0.184 \\
      V407Cyg & 0.976 \\
      Sco2012 & 1.0 \\
      TPyx & 1.0 \\
      SSCyg & 1.0 \\
      V1723Aql & 1.0 \\
    \end{tabular}
    &
    \begin{tabular}[t]{l|c}
      \hline
      \textbf{Name} & \textbf{Acc} \\
      \hline
      GRB030329 & 0.199 \\
      GRB970508 & 0.952 \\
      GRB060418 & 0.983 \\
      GRB110709B & 0.988 \\
      
    \end{tabular}
  \end{tabular}
  \label{tab:drop}
\end{table*}

\autoref{tab:drop} shows the results of the individual dropout test. Samples from one light curve was removed during training. These samples were then used to test the  classifier. The accuracy for the samples of each of the light curves for the four classes are shown. From left to right the classes are: AGN, XRB, SN, Nova and GRB. It can be seen that the classifier is very accurate for AGNs, XRBs and GRBs but misclassifies SNe and Novae. This shows that the intrinsic variabilities of these two classes are not captured by our dataset.


\section{Flux features}
\label{sec:simple_res}



\subsection{Flux feature extraction}
\label{sec:simple}


\mln{This section details an investigation of a set of features that were ultimately not used in this paper. We include it here because the features, a simple difference in flux over different time steps, are likely to be an obvious choice for this problem and we thus consider it useful to demonstrate that they are likely inadequate for transient classification.} 

The feature vector, $\Delta F$, was defined to be

\begin{equation}
  \Delta F = F_{t} - F_{t_0}\;,
\end{equation}

\ksn{where $\Delta F$ is the difference in flux between a reference flux, $F_{t_0}$, chosen at random from the light curve and the flux, $F_t$, at time $t > t_0$.}

In anticipation of a fast imager on MeerKAT we chose the minimum difference between successive flux measurements to be two seconds. The maximum time difference was chosen to be three months to account for the objects that vary on very long time scales such as AGN. The complete set of $t$ where chosen to be:

\begin{align}
  \label{eq:t}
  t = [&2\;\text{sec}, 1\;\text{min}, 5\;\text{min}, 10\;\text{min}, 30\;\text{min}, 1\;\text{hr}, 2\;\text{hr}, 4\;\text{hr}, 6\;\text{hr}, \nonumber \\
  &8\;\text{hr}, 12\;\text{hr}, 18\;\text{hr}, 1\;\text{day}, 2\;\text{day}, 4\;\text{day}, 1\;\text{day}, 2\;\text{week}, \nonumber \\
  &3\;\text{week}, 1\;\text{month}, 1.5\;\text{month}, 2\;\text{month}, 3\;\text{month} ]
\end{align}

The feature extraction method was then as follows. \ksn{First,} GP regression was \ksn{performed} on the original data set. A reference time, $t_0$, was then drawn at random to be somewhere within the curve. The GP was then sampled at this $t_0$ to obtain an $F_{t_0}$. The GP was then sampled at points $(t_0 + t)$. $F_{t_0}$ was then subtracted from these fluxes to obtain the feature vector $\Delta F$. This process was \ksn{repeated} multiple times for each light curve, each time generating a random $t_0$. This was done to simulate the fact that the transient may be detected at any point on the light curve.

\subsection{Results}

\begin{figure}
  \centering
  \includegraphics[width = \columnwidth]{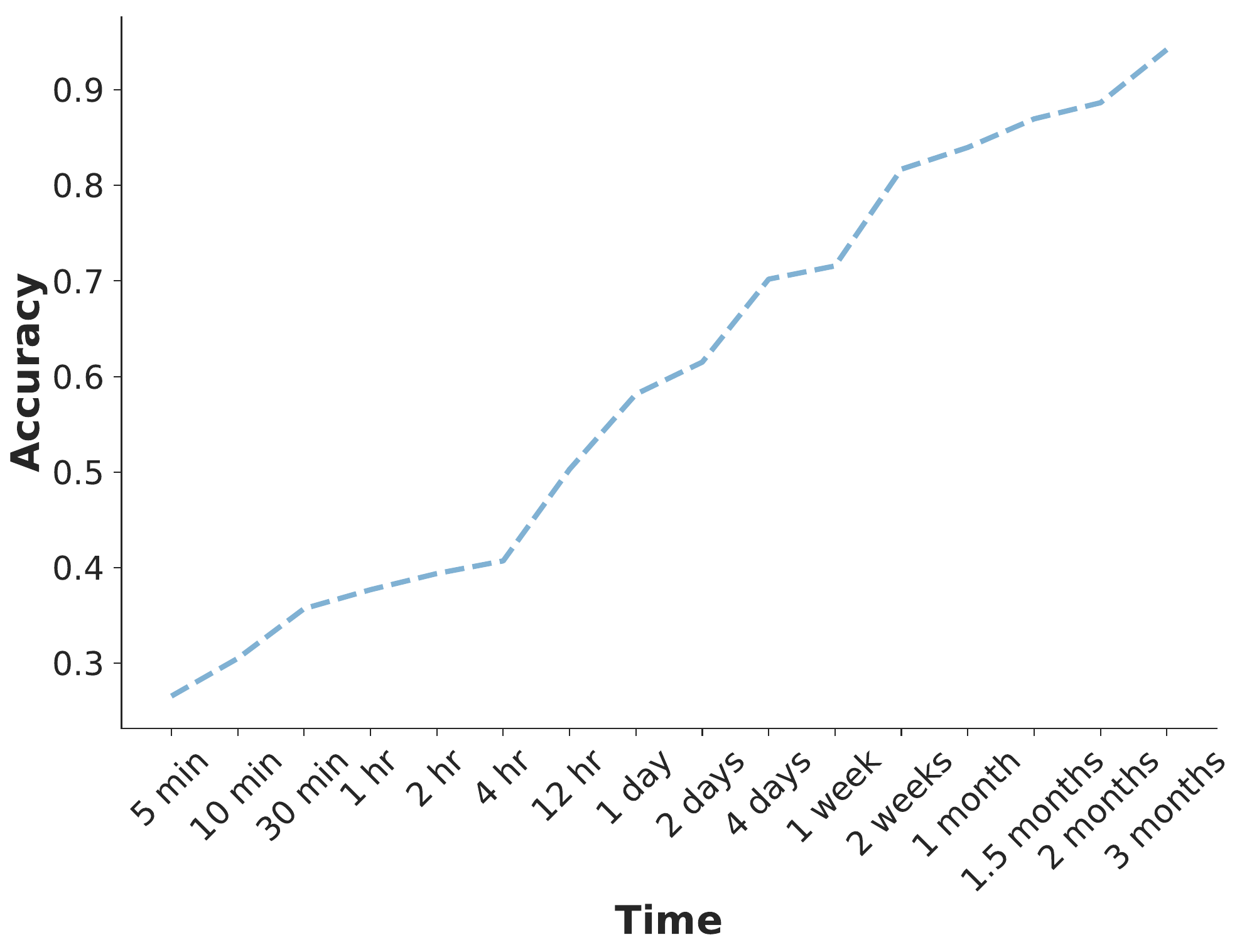}
  \caption{The accuracy of different random forest classifiers each trained on increasing timescales of the total feature set. The $y$-axis shows the accuracy and the $x$-axis shows the time observed. It can be seen that as we increase the time observed the accuracy of the classifier also increases.}
  \label{fig:accVtime}
\end{figure}

Once the flux features were extracted as described in \textcolor{red}{Sec.} \ref{sec:simple}, different subsets of the total feature set was used to train different classifiers. The subsets used were created by truncating the features at different timescales, i.e a classifier was trained on all features  up to a maximum of $5 \text{ min}$, then another was trained on all features  up to a maximum of $ 10 \text{ min}$ and so on for all $t$ in \textcolor{red}{Eq.} \ref{eq:t}. The accuracy of each of these classifiers is shown in \textcolor{red}{Fig.} \ref{fig:accVtime}.

It can see from \textcolor{red}{Fig.} \ref{fig:accVtime} that the classifier performs well on long timescales. \ksn{We would like to investigate is how well the classifier performs on short time scales for each of this classes. It can be seen from \textcolor{red}{Fig.} \ref{fig:numVtime} that the number of FS light curves in the dataset goes to zero, hence 8 hrs is the longest timescale at which we have a complete set of classes. Thus the time vector from \autoref{sec:simple} is changed to:} 

\begin{equation*}
  t = [2\;\text{sec}, 1\;\text{min}, 5\;\text{min}, 10\;\text{min}, 30\;\text{min}, 1\;\text{hr}, 2\;\text{hr}, 4\;\text{hr}, 6\;\text{hr}, 8\;\text{hr}]
\end{equation*}

\begin{figure*}
  \begin{subfigure}{.5\linewidth}
    \centering
    \includegraphics[width = \columnwidth]{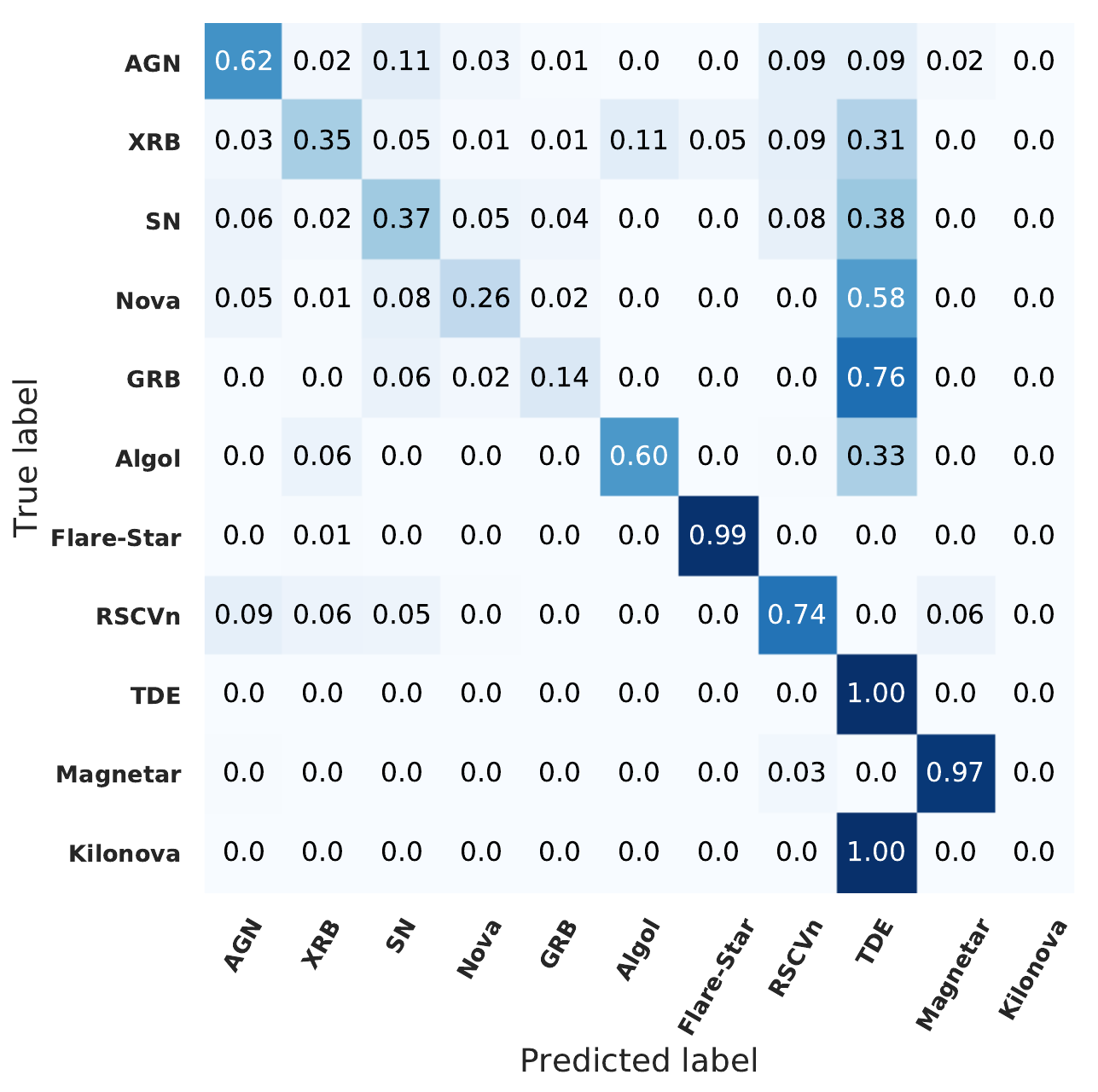}
    \caption{Confusion matrix without contextual information}
  \end{subfigure}%
  \begin{subfigure}{.5\linewidth}
    \centering
    \includegraphics[width = \columnwidth]{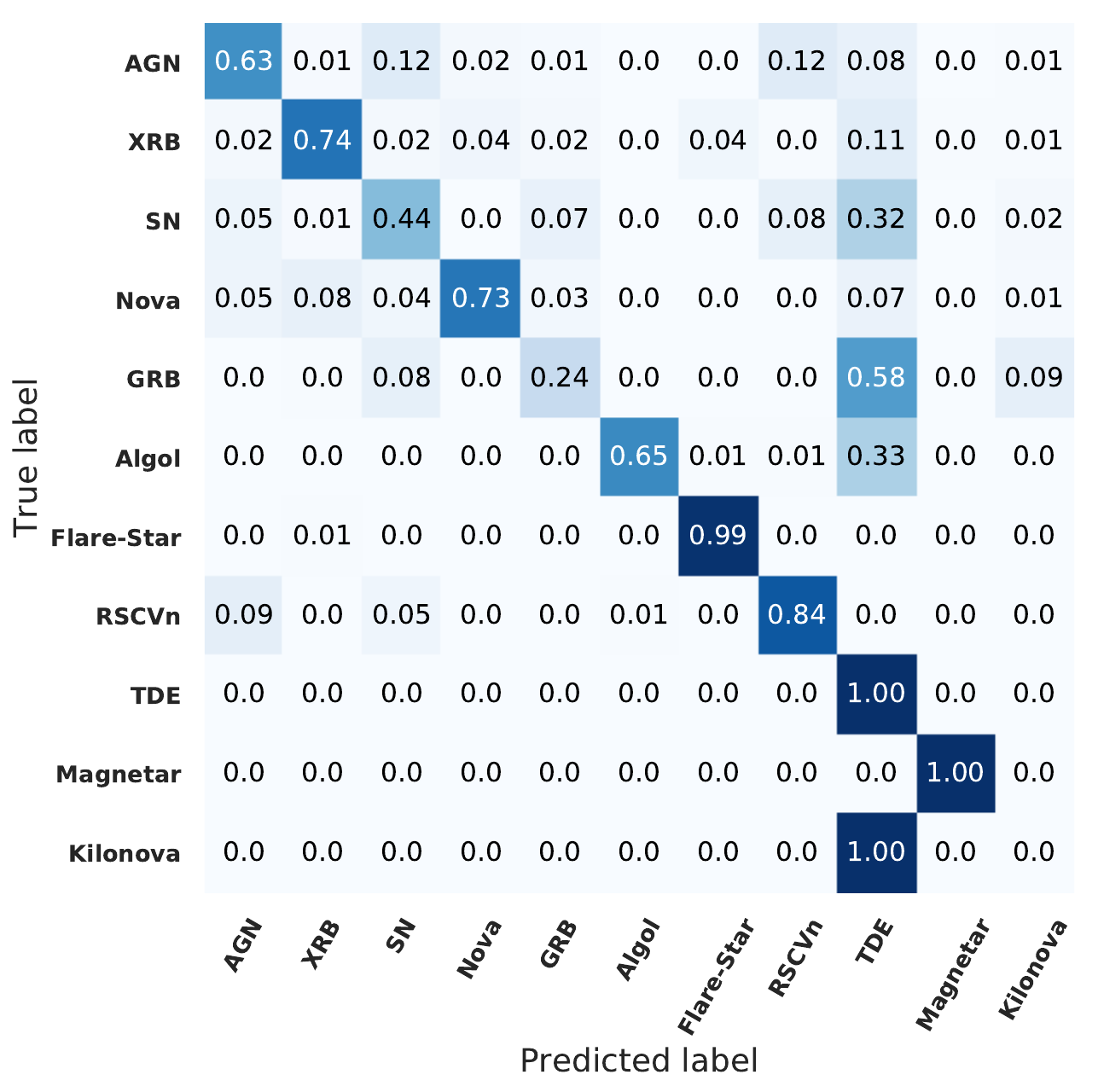}
    \caption{Confusion matrix with contextual information}
  \end{subfigure}
  \caption{The normalised confusion matrix for flux features extracted from 8 hrs of data. The y-axis shows the true label of the object (true class). The x-axis shows the label which the algorithm predicts for the object (predicted class). It can be seen that while contextual information does improve the performance of the classifier this method of feature extraction does not do well to separate the classes.}
  \label{fig:cm_f_8}
\end{figure*}


\section{Variable and transient \ml{confusion} matrices}

\begin{figure}
\centering
\includegraphics[width = \columnwidth]{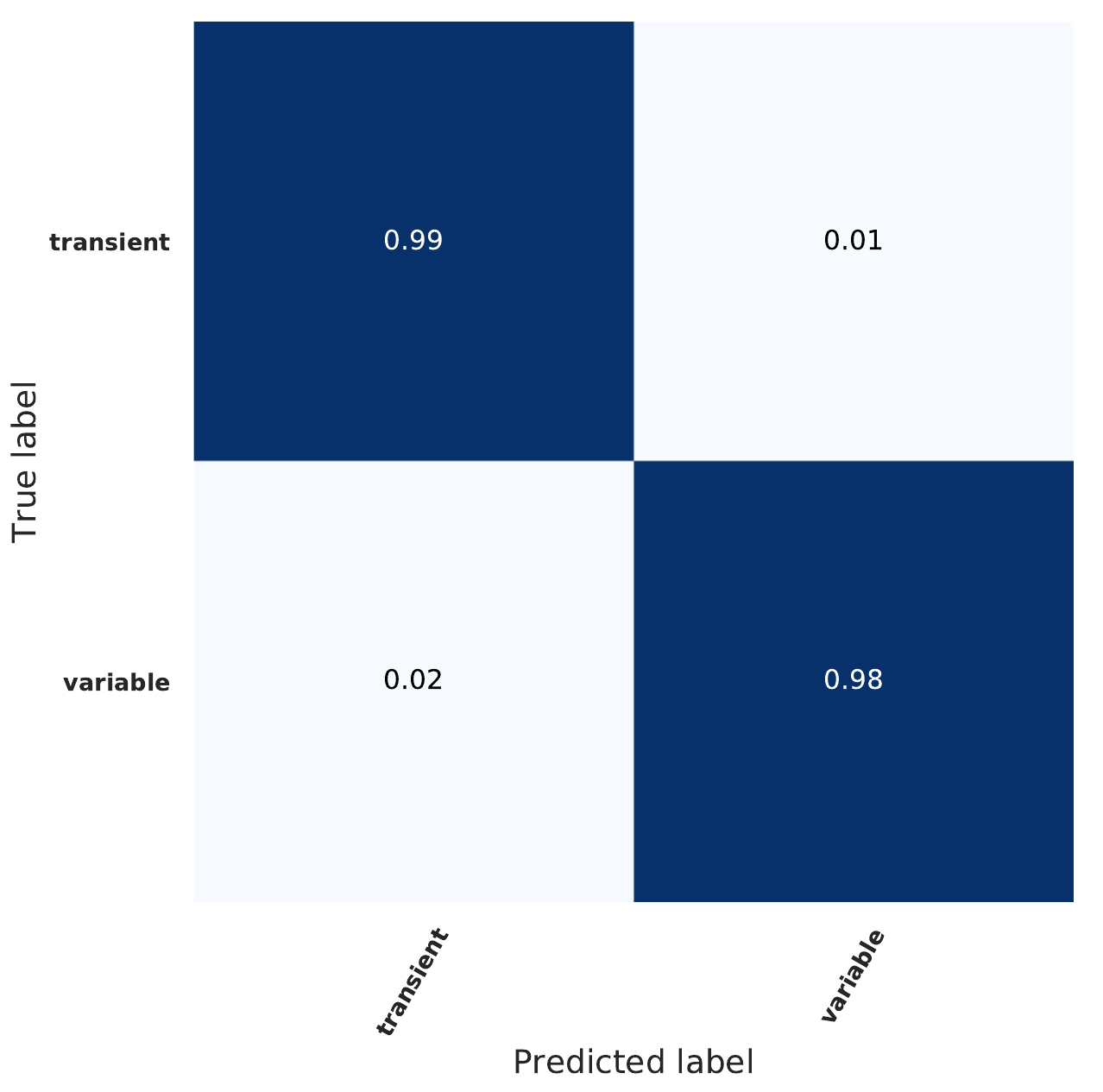}
\caption{\ksn{The normalised confusion matrix for wavelet features extracted from 8 hrs of data. The y-axis shows the true label of the object(true class). The x-axis shows the label which the algorithm predicts for the object(predicted class). It can be seen that the classifier can distinguish between variables and transients extremely well.}}
\label{fig:bin}
\end{figure}
\begin{figure*}
\begin{subfigure}{.5\linewidth}
\centering
\includegraphics[width = \columnwidth]{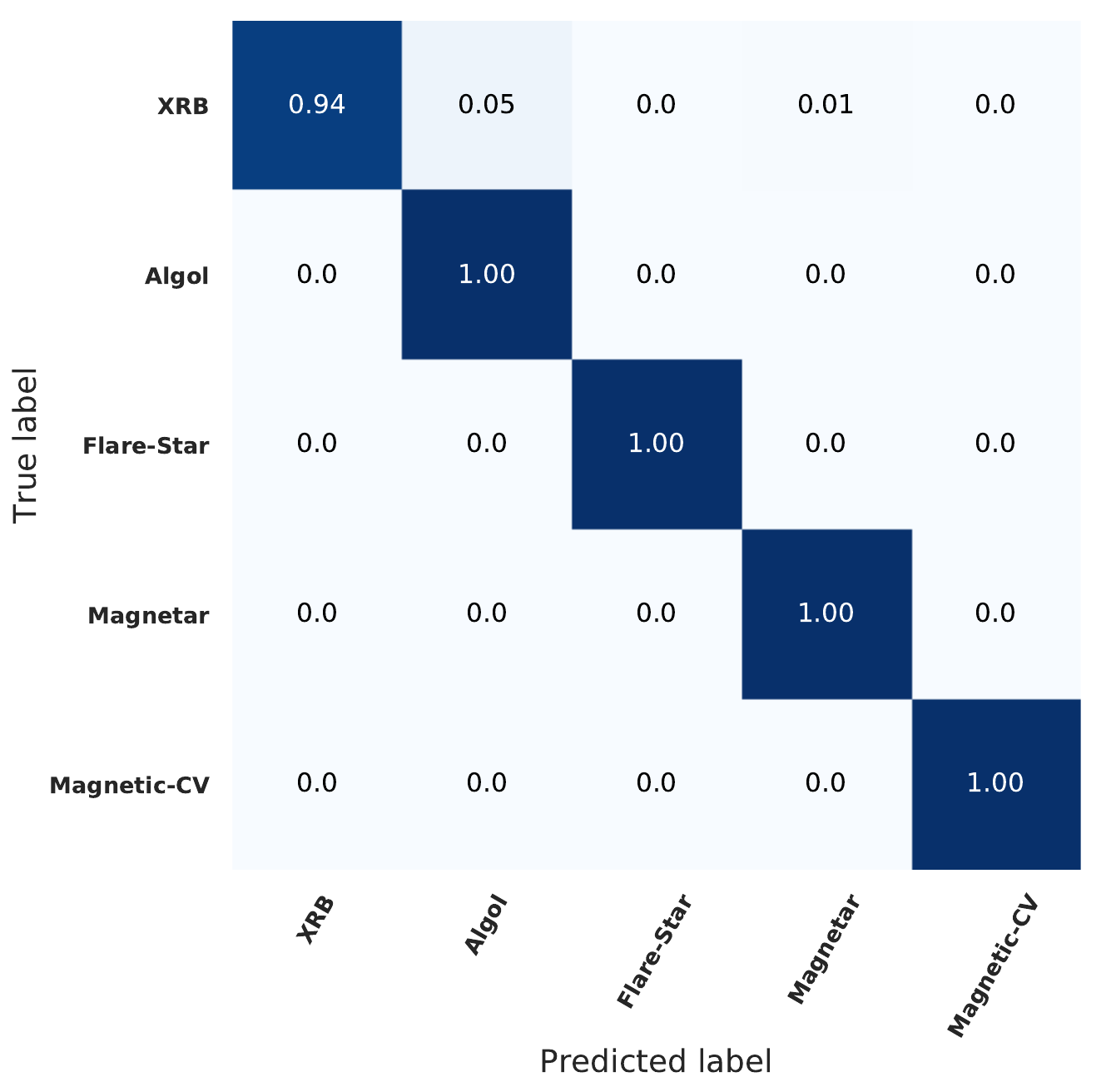}
\caption{Confusion matrix for variable classification}
\end{subfigure}%
\begin{subfigure}{.5\linewidth}
\centering
\includegraphics[width = \columnwidth]{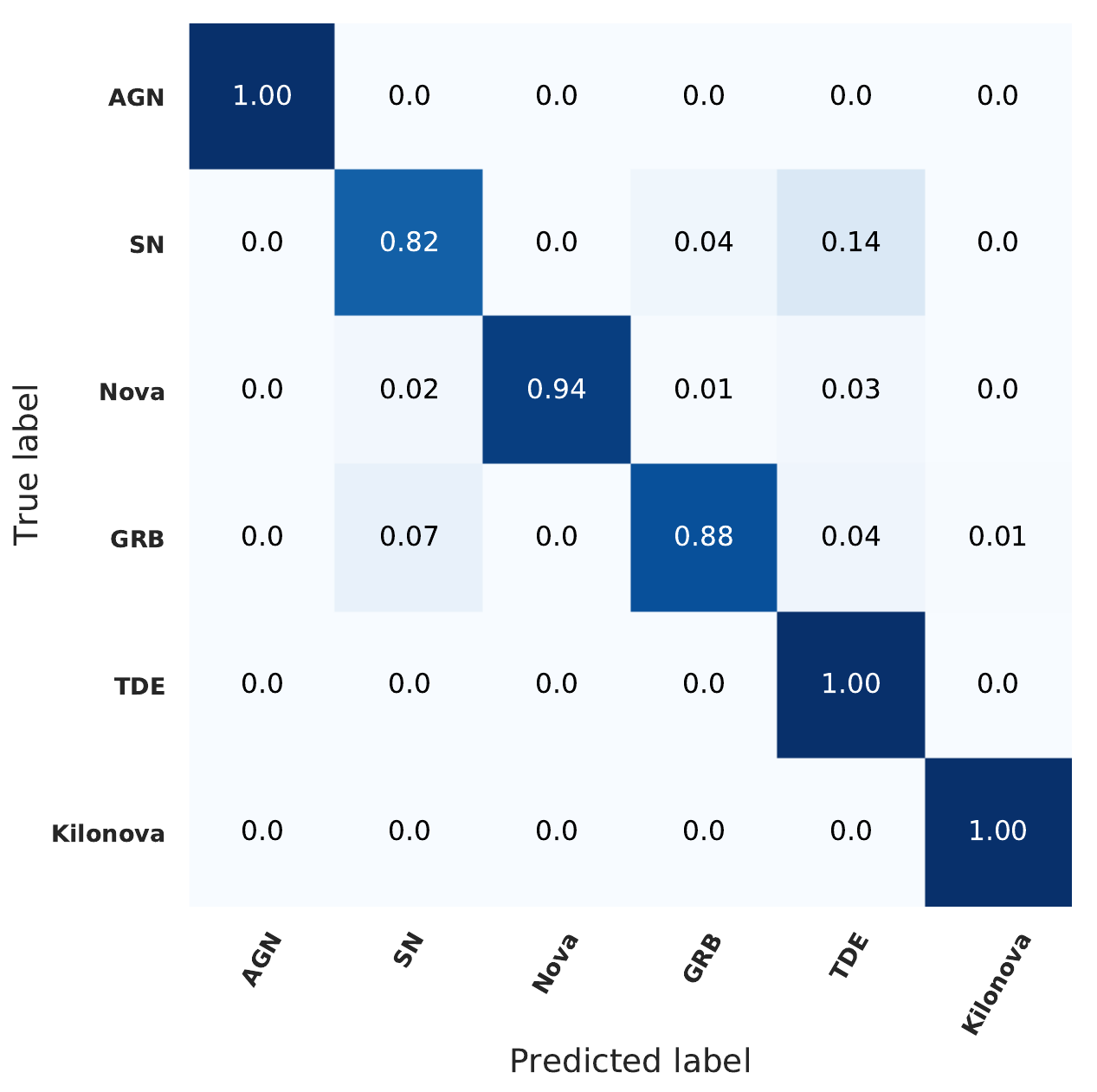}
\caption{Confusion matrix averaged over all runs}
\end{subfigure}
\caption{The normalised confusion matrix for wavelet features extracted from 8 hrs of data. The y-axis shows the true label of the object(true class). The x-axis shows the label which the algorithm predicts for the object(predicted class). By making this split we increase the accuracy of the XRB and SN classifications. The class of GRB, however, sees no improvement. As with before SNe are still being confused with TDEs. It can be seen that this split is unnecessary as it achieves little to no improvement in the performance of the classifier.}
\label{fig:bin_class}
\end{figure*}

\ksn{From \textcolor{red}{Fig.} \ref{fig:data} it can be seen that radio transients can be split into two types: transient that have multiple outbursts on short time scales (e.g. FS) and transients that have one outburst that then decays over longer time scales (e.g. SNe). These two groups are called variables and transients respectively. The dataset was split into these two groups. A classifier was trained on the binary classification of transient vs variable.  Classifiers were then trained on the individual groups of variables and transients. The results of this is shown in \textcolor{red}{Fig.} \ref{fig:bin_class} and \textcolor{red}{Fig.} \ref{fig:bin}}

It can be seen from \textcolor{red}{Fig.} \ref{fig:bin} that the classifier can distinguish between variables and transients extremely well\ksn{, with an overall accuracy of 98.5\%}. From \textcolor{red}{Fig.} \ref{fig:bin_class} we can see that by making this split we increase the accuracy of the XRB and SN classifications \ml{slightly} \ksn{when compared to \textcolor{red}{Fig.} \ref{fig:cm_w_8}}. The class of GRB, however, sees no improvement. As with \textcolor{red}{Fig.} \ref{fig:cm_w_8}, SNe are still being confused with TDEs. It can be seen that this split is unnecessary as it achieves little to no improvement in the performance of the classifier.


\section{Investigation on augmentation.}
\label{app:aug}

\mln{In this appendix, we investigate the value of the different components of our augmention procedure based on Gaussian processes. We first redo our non-representative analysis, where 25\% of the light curves from each class are entirely excluded from the training set. This time however, we only draw \emph{one} 8 hr sample from each light curve, instead of many. We then explore the impact of adding more examples from the long light curves, balancing the data and finally sampling from the Gaussian process to fully understand the effect of augmentation. We note that because we have few light curves, the variance in each of our tests (which we repeat 100 times) is large. Thus we find that while data augmentation essentially always improves performance, the effect it has dramatically depends on the exact training data used. }

\begin{figure}
\centering
\includegraphics[width = \columnwidth]{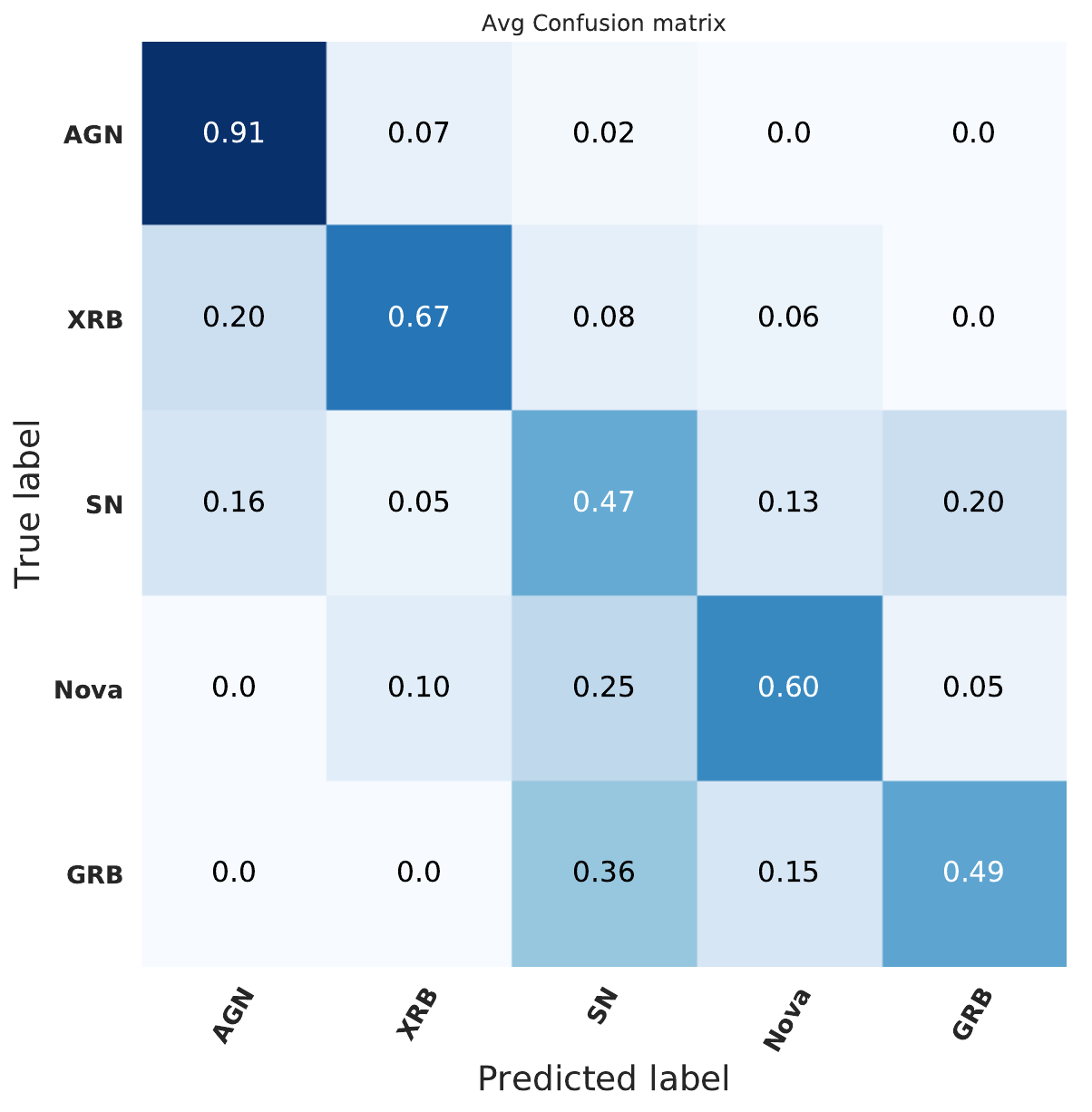}
\caption{\ksd{The normalised confusion matrix averaged over 100 runs. The y-axis shows the true label of the object(true class). The x-axis shows the label which the algorithm predicts for the object(predicted class). The classifier performance is decreased in comparison to Fig. \ref{fig:real}, especially for classes with few examples. }}
\label{fig:cm_unbal}
\end{figure}

\ksd{We restricted our dataset to the five main classes used in \autoref{sec:results}. Thus our dataset contained 30 AGN, 14 XRBs, 13 SNe, 8 Novae and 4 GRBs. GP regression was performed on these and one 8 hr observation was sampled from each of the light curves resulting in an unbalanced dataset containing a total of 69 light curves. A classier was then trained on this dataset. The results are shown in  \textcolor{red}{Fig.}  \ref{fig:cm_unbal}.}

\ksd{Fig. \ref{fig:cm_unbal} shows degraded performance in comparison to the original Fig. \ref{fig:real}. Because we leave out the additional classes in this case, we could actually expect performance to be even worse. It does show however that AGN are still very straightforward to classify owing to their lack of variation over the timescale we are interested in. We argue that augmentation can improve this base performance, particularly for classes with few examples. We test this by performing three experiments using different methods to increase the number of light curves in our dataset.
}

\ksd{The first experiment investigates the effect of splitting the light curves up into many 8 hr `chunks' to increase the number of light curves. In the second experiment we increase the number of light curves while also balancing the number of objects per class. Finally, we investigate the effect of sampling from the GPs to add additional variance to the training data. The method of generating the test set was kept constant for all three experiments (the same as Sec. \autoref{sec:drop20}). In addition, the exact same test objects were used for each set of augmented training data, ensuring that the accuracy could be directly compared between datasets.}

\subsection{The effect of splitting light curves.}

\ksd{The radio light curves from the five main classes mentioned above are all relatively long; therefore we could increase the training set size by splitting the light curves into 8 hr pieces. The GPs were used as a statistically rigorous method for interpolation rather than a method for augmentation, because we sample these `chunks' from the mean function. We extract 8 hr chunks in sets of 1, 2, 5, 10, 30 and 60 per light curve, effectively increasing the training data each time. It is important to note here that the original ratio of the class numbers is the same i.e each dataset is still imbalanced with the total number of light curves increased by 2x, 5x etc.} 

\ksd{A classifier was trained on the each of the sampled datasets. This was then repeated 100 times each time randomising the training and test sets as described in \autoref{sec:drop20}. It is important to note that the test set was balanced, consisting of around 200 samples per class drawn from the test light curves. It is only the training data that is imbalanced. The percentage increase in accuracy was then calculated for each of the datasets with respect to the dataset containing one sampled light curve per original light curve. This easily shows whether or not there is an increase in performance independent of the intrinsic accuracy between runs. }

\begin{figure*}
\begin{subfigure}{.5\linewidth}
\centering
\includegraphics[width = \columnwidth]{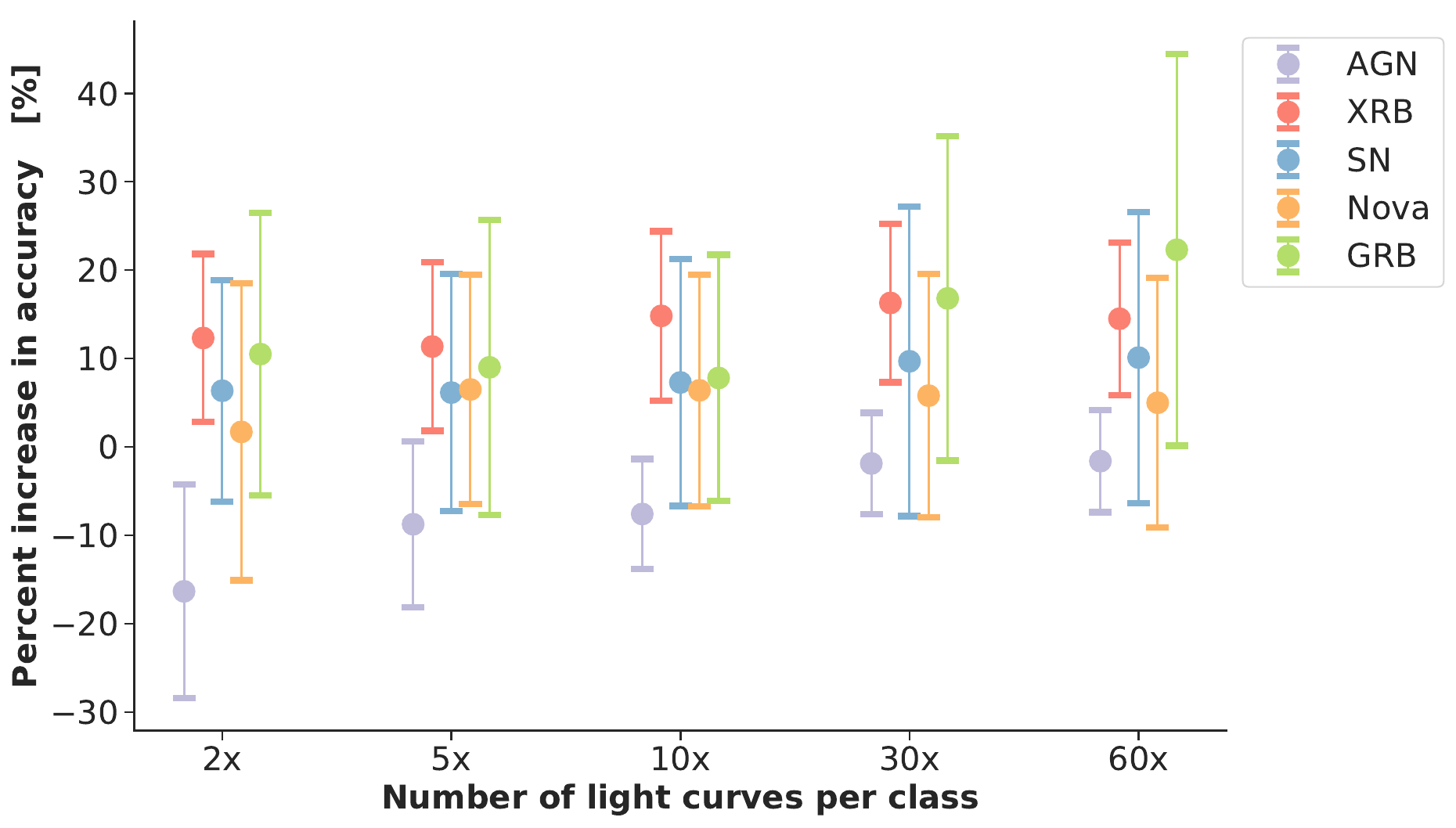}
\caption{The average increase in accuracy per class.}
\end{subfigure}%
\begin{subfigure}{.5\linewidth}
\centering
\includegraphics[width = \columnwidth]{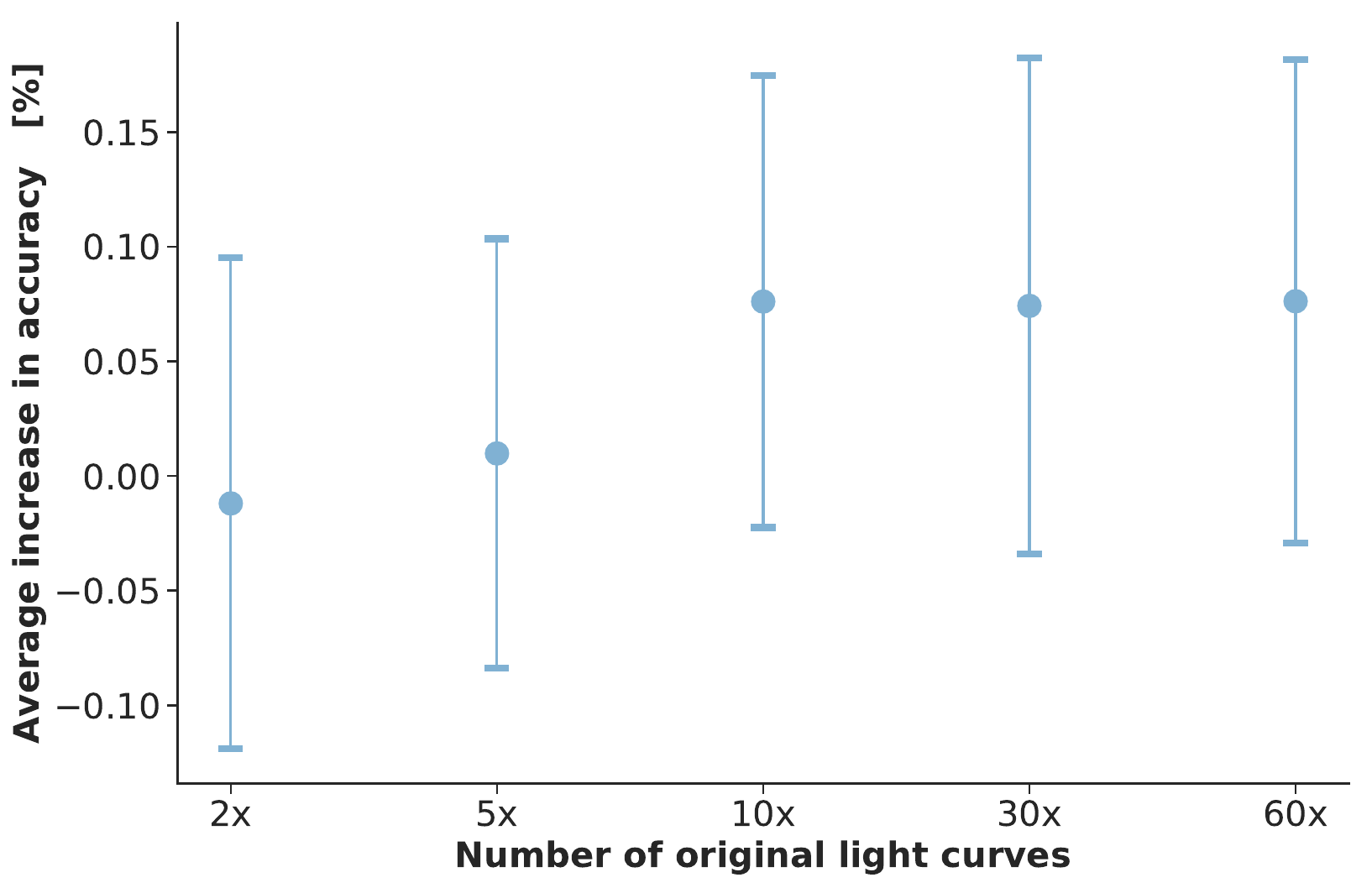}
\caption{The average increase in overall accuracy}
\end{subfigure}
\caption{\ksd{The increase in accuracy per class and the overall increase in accuracy as a function of dataset size relative to the original dataset. We can see that initially the performance of the classifier improves on all classes  except for AGN. AGN are the majority class thus the classifier learns that it is beneficial to label more objects as AGN. As we increase the number of objects in the smaller classes the classifier learns to better classify them leading to less objects being mis-classified as AGN.}}
\label{fig:avg_unbal}
\end{figure*}

\ksd{The average percentage increase in accuracy is shown in \textcolor{red}{Fig.} \ref{fig:avg_unbal}. From this figure it can be seen that the accuracy of all classes except for AGN increase. AGNs are the class with the highest number of objects, thus  the algorithm learns that it is  beneficial to label more objects as AGNs hence AGNs have a high accuracy. As we increase the number of objects in the smaller classes the classifier learns to better classify them, thus less objects are mis-classified as AGN and hence the accuracy of AGNs decrease. \textcolor{red}{Fig.} \ref{fig:avg_unbal} also shows the general trend that increasing the number of light curves in the dataset leads to an increase in overall accuracy.}


\subsection{The effect of increasing the number of light curves while balancing the classes using GPs}

\begin{figure}
\centering
\includegraphics[width = \columnwidth]{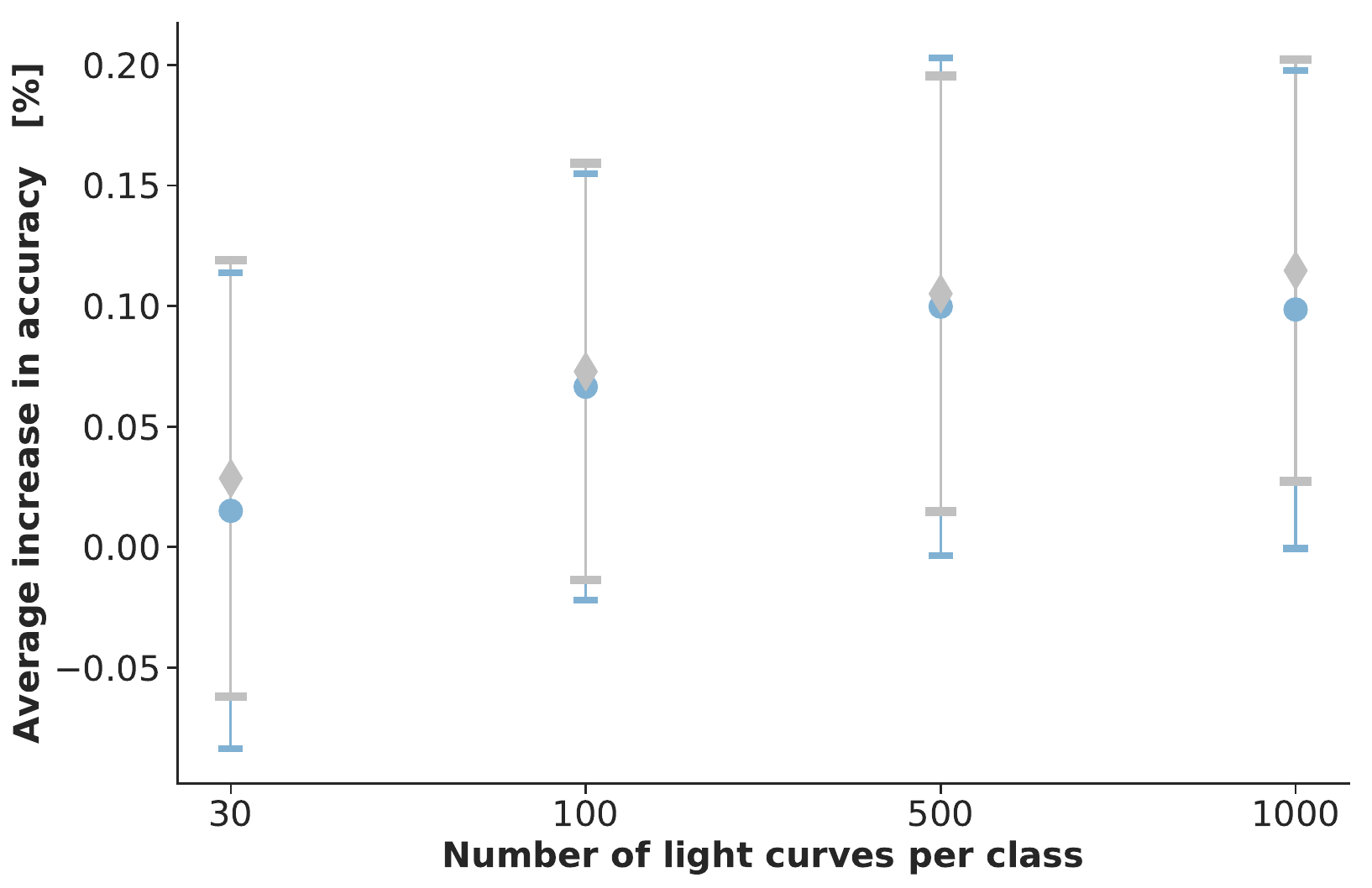}
\caption{\ksd{The average increase in overall accuracy of the classifier for different number of light curves. The blue points show the performance of classifiers trained on light curves sampled from the mean of the GPs. The grey points show the performance of classifiers trained on light curves sampled using the mean as well as the variance of the GPs. We can see that increasing the number of light curves while balancing the classes substantially improves the performance of the classifier, increasing the overall accuracy by $\approx 10\%$. Using the variance of the GPs further improves the overall accuracy by $\approx 2\%$ for any number of extra light curves.}}
\label{fig:avg_bal}
\end{figure}

\ksd{The experiment proceeded as follows: GP regression was performed on each of the original light curves as before. We then sampled 8 hr chunks from the means of these GPs this time creating balanced datasets, where each class contained 30, 100, 500 and 1000 light curves respectively. As before, a classifier was trained 100 times on the each of the sampled datasets, randomising the training and test sets as described in \autoref{sec:drop20}. The percentage increase in accuracy was then calculated for each of the datasets with respect to the dataset containing one sampled light curve per original light curve. \textcolor{red}{Figure.}  \ref{fig:avg_bal} shows the overall increase in accuracy over the 100 runs plotted in blue.} 

\ksd{Finally, we explore the impact of adding variation by drawing samples from the GP instead of just using the mean. These are plotted in grey in \textcolor{red}{Fig.} \ref{fig:avg_bal}. From these we can see that both splitting up the light curves using the means of the GPs as well as sampling from the entire GP improves the performance of the classifier. Using random samples drawn from the GP instead of just the mean improves the accuracy by $\approx 2\%$ on average and slightly reduces the variance in the accuracy between runs. Comparing this to the previous experiment we see that for similar sized datasets, the balanced dataset outperforms the unbalanced dataset by $3 - 5\%$. For this particular problem, augmentation increase accuracy by around $11\%$ on average and could improve performance by considerably more depending on the exact training set used. Thus we recommend the use of augmentation when classifying small, diverse populations of transients.}


\kss{\section{ $t$-SNE- An Alternative Visualisation Tool}}

We ran $t$-SNE for the data in Fig. \ref{fig:umap}, a Euclidean metric was used when estimating the manifold. We found that while the plots are structurally different, they demonstrate the same conclusion.

\begin{figure*}
\begin{subfigure}{.5\linewidth}
\centering
\includegraphics[width = \columnwidth]{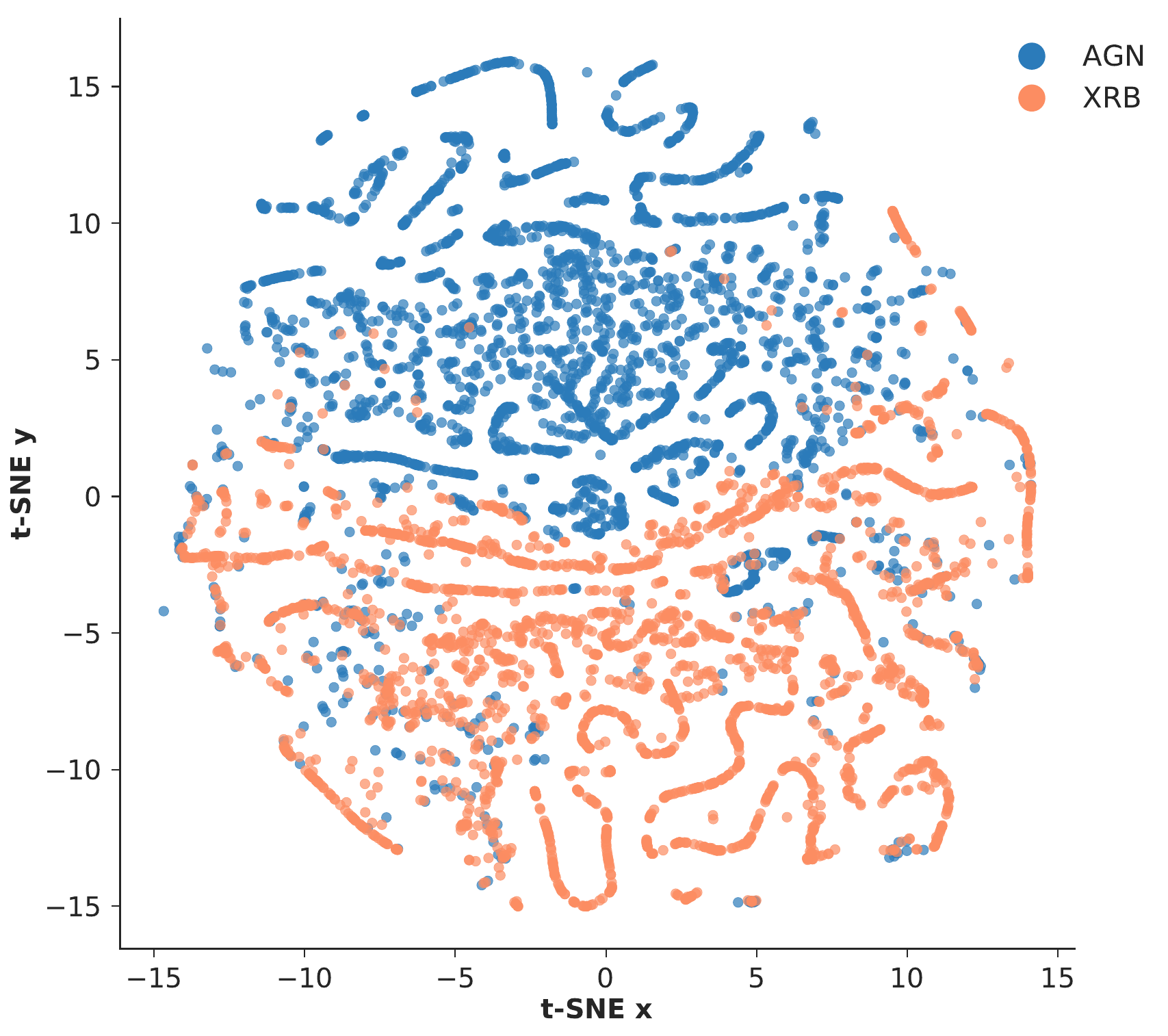}
\caption{AGN-XRB t-SNE}
\label{fig:tsne_agn}
\end{subfigure}%
\begin{subfigure}{.5\linewidth}
\centering
\includegraphics[width = \columnwidth]{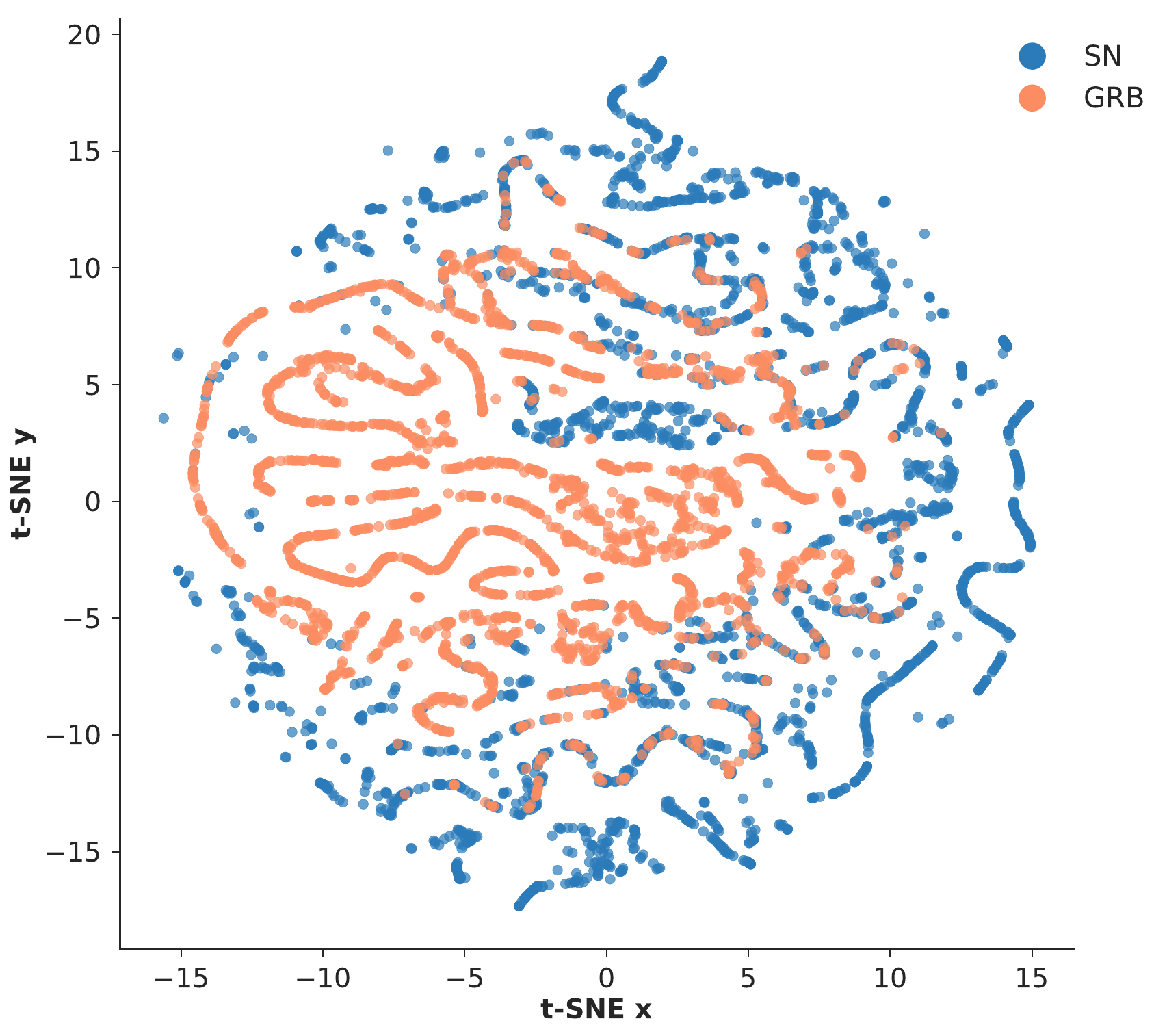}
\caption{SN-GRB t-SNE}
\label{fig:tsne_xrb}
\end{subfigure}\\[1ex]
\caption{An alternative to Figure \ref{fig:umap}, $t$-SNE plots for the four classes of radio transients are shown. $t$-SNE plots show the \ml{high dimensional feature space embedded in} two dimensions. As in Fig. \ref{fig:umap}, the plots show good separation between AGN and XRBs but much more overlap between SNe and GRBs, which are more difficult to classify.}
\label{fig:tsne}
\end{figure*}


\section*{Data availability}

The data underlying this article are available in the Zenodo Digital Repository, at https://dx.doi.org/10.5281/zenodo.4035188




\bibliographystyle{mnras}
\bibliography{ref} 

%

\label{lastpage}
\end{document}